\documentclass{aa}

\usepackage{natbib}
\usepackage{graphicx}
\usepackage{txfonts}
\usepackage{epsfig}
\usepackage{times}
\usepackage{rotating}
\usepackage{float}
\usepackage{setspace}
\usepackage{lscape}
\usepackage{epstopdf}
\usepackage{tabularx}
\usepackage{graphicx} 
\usepackage{subfigure}

\begin{document}

   \title{Photometric redshifts for the Pan-STARRS1 survey}

   \author{P. Tarr\'io
		   	\inst{1,2}
          \and
          S. Zarattini
          \inst{1}
          }

   \institute{IRFU, CEA, Université Paris-Saclay, F-91191 Gif-sur-Yvette, France \\
   	Université Paris Diderot, AIM, Sorbonne Paris Cité, CEA, CNRS, F-91191, Gif-sur-Yvette, France
   	\and
   	Observatorio Astron\'omico Nacional (OAN-IGN), C/ Alfonso XII 3, E-28014, Madrid, Spain\\
   	\email{p.tarrio@oan.es}
   }

   \abstract{We present a robust method to estimate the redshift of galaxies using Pan-STARRS1 photometric data. Our method is an adaptation of the one proposed by \citet{Beck2016} for the SDSS Data Release 12. It uses a training set of 2\,313\,724 galaxies for which the spectroscopic redshift is obtained from SDSS, and magnitudes and colours are obtained from the Pan-STARRS1 Data Release 2 survey. The photometric redshift of a galaxy is then estimated by means of a local linear regression in a 5-dimensional magnitude and colour space. Our method achieves an average bias of $\overline{\Delta z_{\rm norm}}=-2.01 \times 10^{-4}$, a standard deviation of $\sigma(\Delta z_{\rm norm})=0.0298$, and an outlier rate of $P_o=4.32\%$ when cross-validating on the training set. Even though the relation between each of the Pan-STARRS1 colours and the spectroscopic redshifts is noisier than for SDSS colours, the results obtained by our method are very close to those yielded by SDSS data. The proposed method has the additional advantage of allowing the estimation of photometric redshifts on a larger portion of the sky ($\sim 3/4$ vs $\sim 1/3$). The training set and the code implementing this method are publicly available at www.testaddress.com.}


   \maketitle
%

\section{Introduction}\label{sec:intro}

In the last two decades, there has been a rise in the development of large photometric surveys, like the Sloan Digital Sky Survey \citep[SDSS,][]{York2000}, the Panoramic Survey Telescope \& Rapid Response System \citep[Pan-STARRS,][]{Chambers2016}, and the Dark Energy Survey \citep[DES,][]{DES2016}. Robust methods to estimate the redshift of galaxies from photometric data are essential to maximize the scientific exploitation of these surveys. 

Two main approaches are generally used for the computation of photometric redshifts: methods based on physical models and data-driven methods. In the model-based approach, the estimation of the redshift is obtained by modelling the physical processes that drive the light emission of the object. The simplest and most-commonly used method belonging to this category is spectral energy distribution (SED) fitting. It is based on the definition of a SED model, either from theory or from observations, and the fitting of this model to a series of observations in different bands. The definition of an appropriate model is crucial for the performance of the method, therefore it requires taking many different aspects into account (stellar populations models, nebular emissions, and dust attenuation amongst others). Once the model is defined, observations over the entire wavelength range are required for obtaining an accurate fitting. Examples of these methods are the HYPERZ code \citep{Bolzonella2000}, the BPZ code \citep{Benitez2000}, the LePhare code \citep{Ilbert2006}, and the EAZY code \citep{Brammer2008}. \cite{Saglia2012} also apply a SED technique to compute the photometric redshifts of galaxies using Pan-STARRS broadband photometry.

In large-area photometric surveys like SDSS, Pan-STARRS, and DES, the number of photometric bands available is relatively small (5 for each of the cited surveys) and they only cover the optical part of the spectrum. Thus, if no ancillary data are available, the SED fitting technique is not very robust in the determination of photometric redshifts. On the other hand, these surveys offer a large number of extragalactic sources, being well suitable for the use of data-driven methods. These methods usually employ a supervised machine learning algorithm to estimate the unknown redshift of a galaxy from broadband photometry. Supervised algorithms require a (large) set of reliable spectroscopic redshifts that are used to learn how redshifts correlate with colours. Some examples of these techniques are ANNz \citep{Collister2004}, ANNz2 \citep{Sadeh2016},
TPZ \citep{Carrasco2013}, GPz \citep{Almosallam2016}, METAPhoR \citep{Cavuoti2017} or the nearest-neighbors color-matching photometric redshift estimator of \cite{Graham2018}.

Another example of a machine learning approach for the computation of photometric redshifts is presented in \citet{Beck2016}:  a large sample of galaxies (about 2 millions) with both photometric and spectroscopic information is used as training set to estimate the redshift of all the galaxies in SDSS Data Release 12 \citep[DR12,][]{Alam2015} using a local linear regression. A similar method was also presented in \citet{Csabai2007} and in earlier SDSS releases. Nowadays, these photometric redshifts from SDSS are widely used in a variety of scientific publications and the robustness of the algorithm is well established.

The goal of this paper is to adapt the \citet{Beck2016} algorithm to compute photometric redshifts using the Pan-STARRS1 (PS1) photometric data, which cover an area that is twice as large as the SDSS footprint and whose magnitude limits are about two magnitudes fainter than the SDSS ones. SDSS and PS1 have four photometric bands in common, plus a fifth band that is different, and which is on the bluer side of the spectrum for SDSS and on the redder side for PS1. To adapt \cite{Beck2016} algorithm to PS1 we thus needed to select the appropriate PS1 data that allow us to compute the redshift, to construct a proper training set, and to reassess the performance of the linear regression algorithm when using this information. The training set and the code implementing the PS1 photometric redshift method presented in this paper has been made available for the community at \url{www.testaddress.com}. 

The method proposed in this paper was initially designed with the purpose of confirming cluster candidates of the ComPRASS catalogue \citep{Tarrio2019}. This all-sky catalogue of galaxy clusters and cluster candidates was validated by careful cross-identification with previously known clusters, especially in the SDSS and SPT footprints. Still, many candidates remain unconfirmed outside these areas. Having information on the photometric redshifts in the PS1 area will enable us to confirm ComPRASS candidates in this region. This information will also facilitate the extension of other scientific studies performed with SDSS photometric data to the area of the sky covered by PS1. Some examples are studies related to the formation and evolution of galaxies, or to the properties of dark energy \citep{Salvato2019}.

The paper is organised as follows: Section \ref{sec:method} summarises the linear regression method and how it is adapted to PS1 data. Section \ref{sec:training} describes the procedures that we put in place to prepare the training set using PS1 photometry and SDSS spectroscopy. Section \ref{sec:results}  evaluates the performance of the proposed redshift estimation method. Section \ref{sec:feature_comparison} presents a comparison with the results obtained using different photometric data from PS1 and SDSS. Section \ref{sec:best_practices} gives some practical notes on the use of the method and the associated dataset. Finally, Sect. \ref{sec:summary} concludes the paper with a summary of the main results.

\section{Method}\label{sec:method}
In this Section we describe the method to estimate the redshift of galaxies from PS1 photometric data. This method is an adaptation of the method used in the SDSS DR12 \citep{Beck2016} and can be used to calculate photometric redshifts for all galaxies in the PS1 footprint ($\sim$ 3/4 of the sky). The method is prepared to work on both Data Release 1 (DR1) and Data Release 2 (DR2), although in this paper we will present the results corresponding to DR2. The performance for DR1 data was also tested, finding no significant differences.

The method is data-driven and uses a training set $\mathcal{T}$ composed of galaxies with known spectroscopic redshift and a set of magnitudes and colours obtained from the PS1 survey.  The redshift of a galaxy is estimated by means of a local linear regression in a $D$-dimensional magnitude and colour space. The rest of this Section summarizes the linear regression algorithm (see also \cite{Beck2016}) and describes in detail how we selected the magnitude-colour space to be used for PS1. We also explain how we deal with the potential problem of missing information.

\subsection{Linear regression algorithm}\label{ssec:linear_regression}
The local linear model establishes that the redshift of a galaxy can be written as a linear combination of $D$ galaxy properties (magnitudes and colours), hereafter features, $x_1, ..., x_{\rm D}$, as follows:
\begin{equation}\label{eq:linear_model}
z = \mathbf{x}^{\rm T} \mathbf{\theta}
,\end{equation}
where $\mathbf{x} = [1, x_1, ..., x_{\rm D}]^{\rm T}$ is the feature vector of the galaxy. The column vector $\mathbf{\theta}$ contains the $D$+1 coefficients of the $D$-dimensional linear regression, with its first element representing a constant offset. The coefficient vector $\mathbf{\theta}$ can be estimated by constructing an over-determined system of $k$ equations using $k$ galaxies of the training set $\mathcal{T}$: $\mathbf{z}_{\rm spec} = \mathbf{X} \mathbf{\theta}$, with ${\mathbf{z}_{\rm spec} = [z^{(1)}, ..., z^{(k)}]^{\rm T}}$ being the spectroscopic redshifts of the $k$ chosen galaxies and $\mathbf{X} = [\mathbf{x}^{(1)}, ..., \mathbf{x}^{(k)}]^{\rm T}$ the corresponding $k$ feature vectors. The least-squares solution of this system is then: 
\begin{equation}\label{eq:theta}
\mathbf{\hat{\theta}} = (\mathbf{X}^{\rm T} \mathbf{X})^{-1}\mathbf{X}^{\rm T} \mathbf{z}_{\rm spec}
.\end{equation}

The error of the photometric redshift can be estimated from the difference between the spectroscopic redshifts of the $k$ galaxies and the corresponding photometric redshifts provided by the regression:
\begin{equation}\label{eq:error}
\delta_{z_{\rm phot}} = \sqrt{\frac{\sum_{k}(\mathbf{z}_{\rm spec} - \mathbf{X} \mathbf{\hat{\theta}})^2}{k}}
.\end{equation}

To apply this method, it is necessary to define how to chose the $k$ training galaxies used to estimate $\theta$, and to define the $D$ features that will characterize each galaxy. 

In our case, the $k$ galaxies are chosen to be the nearest neighbours of the target galaxy in terms of Euclidean distance in the $D$-dimensional space. In particular, we chose $k=100$, as in \citet{Beck2016}. Additionally, in the case that some of these $k$ neighbours have outlying redshifts ($|z_{\rm spec}^{(j)}-\mathbf{x}^{(j)\rm T} \mathbf{\hat{\theta}}| > 3\delta_{z_{\rm phot}}$), we discard them and repeat the computation of $\hat{\theta}$ (Eq. \ref{eq:theta}) using the remaining $l<k$ neighbours. We note that, in some cases, a galaxy can fall outside the $D$-dimensional bounding box of its nearest neighbours. In these cases, Eq. \ref{eq:linear_model} constitutes an extrapolation, so the results may be less reliable. The impact of the extrapolation on the estimated photometric redshift is evaluated in Section \ref{ssec:photometric_errors}. Our code provides a flag to indicate these cases.

\subsection{Feature selection}\label{ssec:feature_selection}
The key point to successfully employ the linear regression algorithm described above with PS1 photometric data is to appropriately select the $D$ features to be used. SDSS and PS1 surveys have both imaged the sky using five broadband filters. Four of these filters ($g$, $r$, $i$, and $z$) are similar in both surveys, although with some minor differences \citep{Tonry2012}. The fifth filter, however, is completely different: SDSS uses the $u$ filter, which covers the bluest part of the measured spectrum (at bluer waveleghts than the $g$ filter), whereas PS1 uses the $y$ filter, which spans the reddest part of the spectrum (at redder wavelengths than the $z$ filter). 
The method defined in \cite{Beck2016} uses the SDSS $r$ magnitude, and the $u-g$, $g-r$, $r-i$, and $i-z$ colours to define the 5-dimensional space in which the linear regression takes place to estimate the redshift. Since the $u$-band is not available in PS1, a natural choice inspired in \cite{Beck2016} is to use the PS1 $r$ magnitude, and the four colours that can be constructed with consecutive magnitudes, i.e., $g-r$, $r-i$, $i-z$, and $z-y$. These are the 5 features that we decided to use in our method. However, it is worth noticing that other combinations of the 5 bands are also possible without significant difference in the results, given that all the photometric information is included.  

PS1 database provides several ways of measuring magnitudes and fluxes of objects in its five photometric bands. We used stack photometry since it provides the best signal-to-noise, according to \citet{Magnier2019}. Then, different photometric measurements are available:

\begin{itemize}
	\item PSF magnitudes: obtained from fitting a predefined PSF form to the detection. These magnitudes are especially relevant for point sources (e.g. stars).
	\item Kron magnitudes: inferred from the growth curve, after determining the Kron radius of the object. These magnitudes are especially relevant for non-point-sources.
	\item Aperture magnitudes: they measure the total count rate for a point source based on integration over an aperture plus an extrapolation involving the PSF. According to the PS1 database documentation, this photometry should not be used for extended sources, so we will not use it in our method.
	\item Fixed-aperture measurements: flux measured within several predefined aperture radii (1.03, 1.76, 3.00, 4.63, and 7.43 arcsec).
\end{itemize}

Kron magnitudes are the most appropriate for extended objects like galaxies, so we have chosen to use the $r$-band Kron magnitude for defining our $r$ feature. 
Regarding the four colour features ($g-r$, $r-i$, $i-z$, $z-y$), we have considered two different approaches: they can be calculated either a) from fixed-aperture fluxes, or b) from Kron magnitudes.

The first approach (aperture colours) computes the four colours within a fixed aperture. To obtain the aperture magnitudes within the most appropriate aperture, we selected for each galaxy the $g$, $r$, $i$, $z$, and $y$ fixed-aperture fluxes corresponding to the closest aperture to the $r$-band Kron radius of the galaxy ({\it rkronrad}). Then, the five selected aperture fluxes are converted into aperture magnitudes.

The full PS1 dataset files available for direct download do not provide the above-mentioned fixed-aperture fluxes, which need to be queried to the database. Instead, they provide Kron magnitudes. As an alternative approach, we evaluated the use of these magnitudes to compute the four colours required by our method. We note that the colours constructed from the Kron magnitudes are not physically motivated, since the five different magnitudes are not measured within the same aperture.  However, we will show that they provide very similar results to the ones obtained when using the fixed-aperture colours defined above, so for convenience, we added this alternative in our code. Unless otherwise stated, the results presented in this paper were obtained with the aperture colours calculated from the fixed-aperture fluxes. We include a comparison between the different approaches in Sect.~\ref{sec:feature_comparison}.

\subsection{Feature computation}\label{ssec:feature_computation}

Before calculating
the five features we need to apply a dereddening correction to the downloaded or calculated magnitudes.
Reddening is produced by the scattering of the light by dust in the interstellar medium and it depends on the position of the object in the sky. Therefore, it has to be corrected in order to obtain magnitudes that are more correlated with redshift. 

We obtained this correction in the following way: 
Firstly, we computed the colour excess E(B-V) for each galaxy using the \citet{Schlegel1998} maps.
Then, we obtained the extinction $A_\lambda$ for the $g$, $r$, $i$, and $z$ bands by multiplying the colour excess by the values presented in table 22 of \citet{Stoughton2002} for the $g$, $r$, $i$, and $z$ SDSS filters respectively, which are very similar to the ones used in PS1. For the $y$ band, which is not present in SDSS, we calculated the extinction using the parametrization of \citet{Fitzpatrick1999} taking the effective $\lambda$ of the $y$ band ($\lambda_{\rm eff}=9620$ \AA) 
presented in table 4 of \citet{Tonry2012}. 

We then applied the dereddening correction ($g = g_{\rm downloaded} - A_g$, and equivalently for the other bands), and we computed the five features ($g-r$, $r-i$, $i-z$, $z-y$, and $r$).
 
Each dimension is then standardized, by removing the mean and dividing by the standard deviation of the training set $\mathcal{T}$. In this way, all the features span similar ranges and, thus, contribute with a similar weight to the linear regression. This feature scaling is a common practice in algorithms which use Euclidean distance, like ours, since otherwise the feature with a larger scale (the magnitude in our case) would dominate the computation of the distance.

We note that the zero-point correction that is usually taken into account in public software for the computation of photometric redshifts does not need to be included in our method. The reason is that our method is not sensitive to the addition of constant terms to the features, since it uses standardized features.

\subsection{Missing features}\label{ssec:missing_features}
 
The PS1 dataset contains galaxies for which one or more magnitudes may not be available, resulting in missing features. The method described above can still be applied to estimate the photometric redshift of these galaxies in several ways. In particular, we have decided to calculate the redshift of such galaxies by using only the available features, i.e. we construct the feature vector $\mathbf{x}$ with $D'<D$ features, both for the target galaxy and the training galaxies, and then use Eqs. \ref{eq:linear_model} and \ref{eq:theta} as before. 
In this way, we will use the subset of the training set $\mathcal{T}$ that has all the five features available ($\mathcal{T}_5$) to calculate the redshift of a galaxy that has the five features. Likewise, when a galaxy is missing one feature, we will also use $\mathcal{T}_5$ as training set, but we will not consider the missing feature in any of the training galaxies.
Another possible approach to follow in the case of a missing feature in the target galaxy could be to use as training set the subset of $\mathcal{T}$ that has the other 4 features available ($\mathcal{T}_{4}$, with $\mathcal{T}_5 \subset \mathcal{T}_{4} \subset \mathcal{T}$). In this paper, we report the results corresponding to the first approach, but we note that the second option produces similar results and is also available in the code.
 
It is worth mentioning that, for simplicity, the standardization of the features is done in any case with the mean and the standard deviation of the subset $\mathcal{T}_5$. We  tested that other reasonable choices (e.g. using, for each feature, the mean and the standard deviation of all the galaxies containing that feature) do not yield any significant difference in the results.

\begin{table*}
	\caption{Parameters downloaded from PS1 and SDSS databases.}
	\label{table:query_output}
	\centering 
	\small
	\begin{tabular}{c c c c}
		\hline
		\noalign{\smallskip}
		Parameter &  Database & Table  & Description \\
		\noalign{\smallskip}
		\hline
		\noalign{\smallskip}
		ObjID &  SDSS & \cite{Beck2016}    &    Object ID in SDSS \\ 
		ra &   SDSS & photoprimary   &    ra in SDSS \\ 
		dec &   SDSS & photoprimary   &    dec in SDSS \\ 
		objid &    PS1 & StackObjectThin &    Object ID of closest object in PS1  \\ 
		distance &  PS1 &  - &    Distance between SDSS and PS1 positions \\  
		ra &    PS1 & StackObjectThin  &    ra in PS1 \\ 
		dec &   PS1 & StackObjectThin   &   dec in PS1  \\ 
		\{g,r,i,z,y\}KronMag &  PS1 &  StackObjectThin   &   Kron magnitudes  \\
		\{g,r,i,z,y\}KronMagErr &  PS1 &  StackObjectThin   &   Kron magnitudes errors  \\
		primaryDetection &   PS1 & StackObjectThin   &   primary stack detection flag  \\
		rKronRad &   PS1 & StackObjectAttributes   &   Kron radius in $r$ band  \\
		\{g,r,i,z,y\}c6flxR\{3,4,5,6,7\} & PS1 &   StackApFlxExGalCon6   &  fluxes within 5 different apertures   \\
		\{g,r,i,z,y\}c6flxErrR\{3,4,5,6,7\} &  PS1 &  StackApFlxExGalCon6   &  errors in aperture fluxes   \\
		\noalign{\smallskip}
		\hline
	\end{tabular}
\end{table*}

\section{Training set}\label{sec:training}

The training set of \citet{Beck2016} included more than 2 millions galaxies with spectroscopic redshifts. In this work, our goal was to use the same training set, but with features obtained from PS1 magnitudes instead of SDSS. To construct it, we made use of the {\tt CasJobs} tool in SDSS, which allows querying both the SDSS and the PS1 databases. In the catalogue of \citet{Beck2016} each galaxy is identified via its {\it ObjID} (a unique number assigned to each object in the SDSS database). Thus, our first step was to obtain the coordinates for each object using an appropriate query to the SDSS database. Then, we performed a query in the PS1 database \citep{Flewelling2019} to look for the PS1 object nearest to each SDSS object. This was done using the {\tt fGetNearestObjEq} function, and limiting the search to a radius of 30\arcsec. 
This conservative choice allowed us to define {\it a posteriori} the matching distance up to which we can consider the match reliable. Objects with greater matching distances are not kept in the training set. We will show later that this maximum distance was fixed to 1\arcsec. 

Our query produces the following output parameters:
the identifier ({\it ObjID}) and coordinates (ra, dec) of the object in both the SDSS and PS1 databases, the distance between the two positions, the $g$, $r$, $i$, $z$, and $y$ Kron magnitudes and their associated errors in the PS1 database (using stack photometry), the PS1 {\it primarydetection} flag, the Kron radius measured in the PS1 $r$ band, and the $g$, $r$, $i$, $z$, and $y$ PS1 fluxes measured within the five predefined aperture radii (1.03, 1.76, 3.00, 4.63, and 7.43 arcsec) and their associated errors. Table \ref{table:query_output} lists these parameters and the tables where they are available. 

The training set $\mathcal{T}$ is constructed from this catalogue, after cleaning it for unwanted objects, calculating the five features defined in Sect. \ref{sec:method}, and taking the spectroscopic redshift from the catalogue of \citet{Beck2016}. In the following Subsections, we describe these steps in detail.

\begin{figure}[]
	\centering
	\includegraphics[width=0.99\columnwidth]{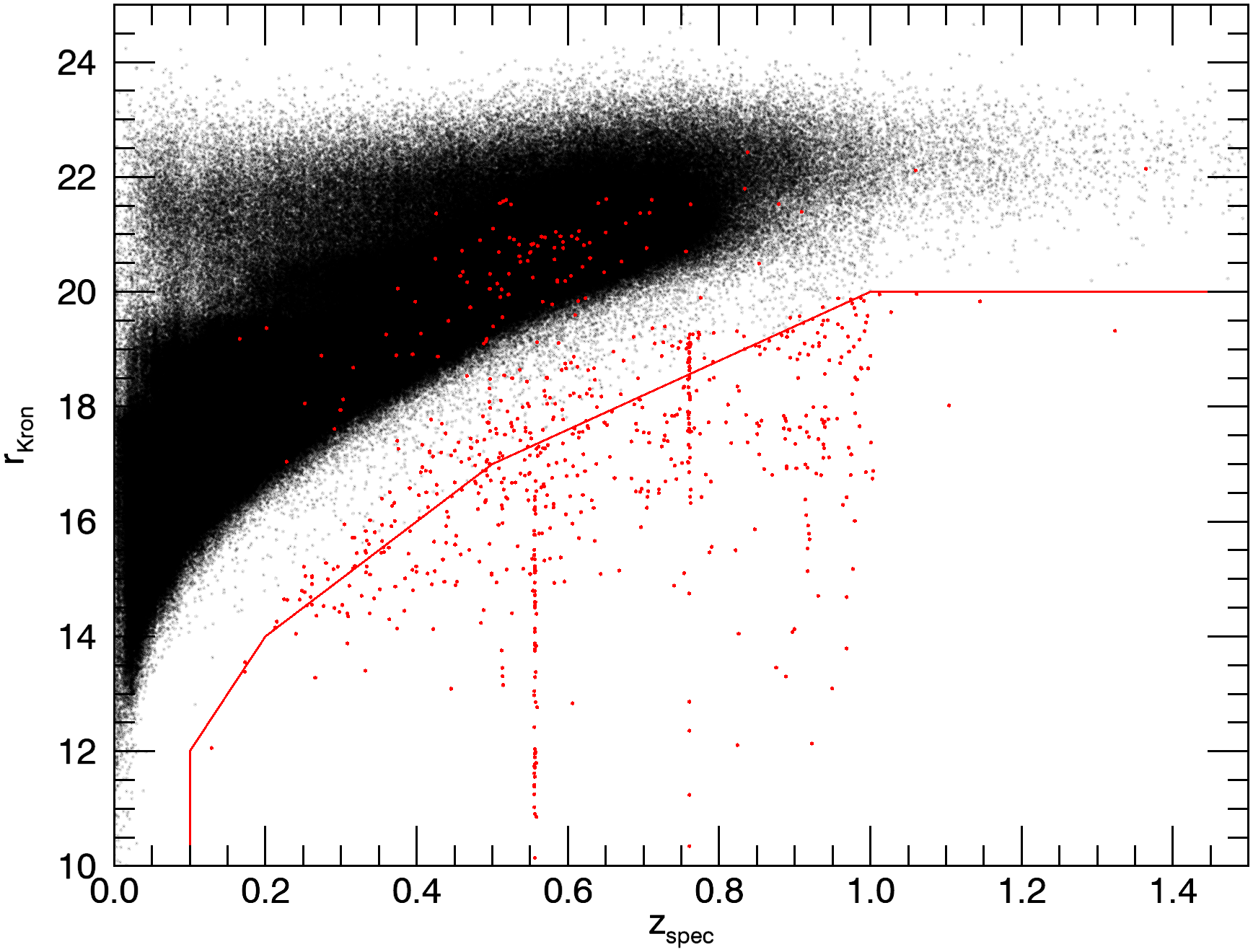}
	\caption{Scatter plot of the $r$ Kron magnitude as a function of the spectroscopic redshift for the galaxies in the training set (black dots). Red dots represent the galaxies that are removed for not satisfying the magnitude limits defined in Table \ref{table:mag_limits}. The red line represents the magnitude limits for the $r$-band.  
	}
	\label{fig:quality_limits}
\end{figure}

\begin{figure}[]
	\centering
	\includegraphics[width=0.99\columnwidth]{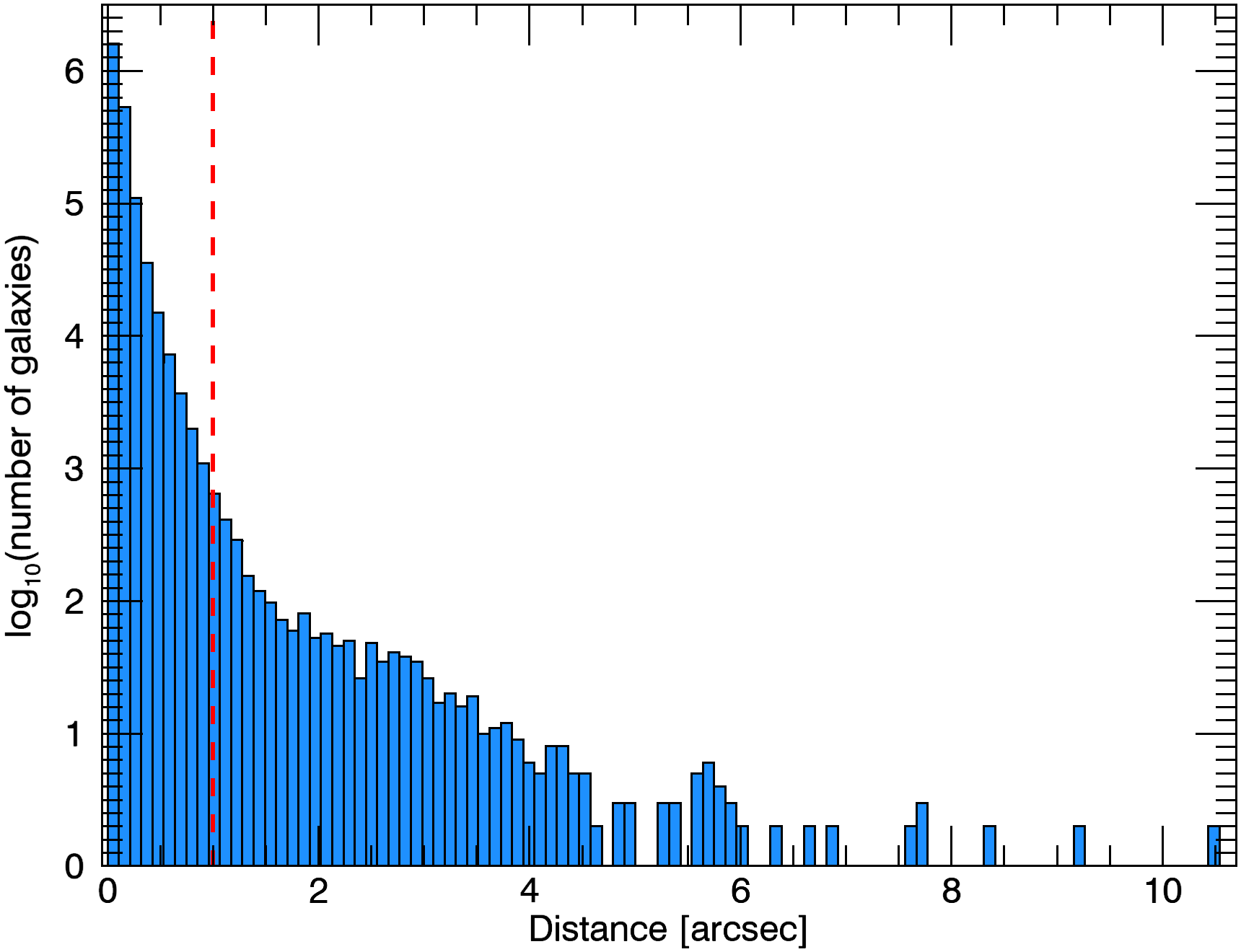}
	\caption{Distribution of the distances between the PS1 objects and the SDSS objects. In the final training sample $\mathcal{T}$ we only used objects with distances smaller than 1\arcsec.}
	\label{fig:distance_distribution}
\end{figure}

\subsection{Cleaning}\label{ssec:cleaning}
The catalogue resulting from the query to the PS1 database contains some duplicate entries, i.e. objects with the same {\it ObjID} and the same or different properties. We cleaned this catalogue from these objects by keeping only one object for each {\it ObjID}. 
In particular, we selected the ones for which {\it primarydetection} was equal to 1, which indicates that the entry is the primary stack detection. If there was more than one object satisfying this condition, we selected the one with more magnitudes available. Exact duplicates were also removed.

We additionally removed from the downloaded catalogue two classes of objects. 
The first class corresponds to objects for which PS1 and SDSS photometry are very different. PS1 and SDSS have four photometric bands in common ($g$, $r$, $i$, and $z$), so we expect to have a small difference between the magnitudes in those four bands measured by PS1 and SDSS. However, we noticed that our catalogue included some objects for which the difference between these magnitudes was very high (even more than ten magnitudes in some cases).  
For precaution, we decided to exclude them from our training set. In particular, we excluded objects for which the difference between any SDSS magnitude and the corresponding PS1 magnitude is greater than 4. 

The second class of excluded objects corresponds to those that appear to be too bright for their assigned spectroscopic redshift. We noticed the presence of very bright objects at high redshift in our catalogue that are not physically possible (e.g. $r=12.1$ at $z=0.82$). After a visual inspection in SDSS we found that these objects were indeed bad samples due to two main reasons: a) low-redshift star-forming galaxies with wrong (high) SDSS spectroscopic redshift; and b) wrong magnitude measurements (for example, in galaxies affected by the light of close-by saturated stars, or by external regions of foreground extended galaxies).  
We decided to remove these objects by setting, for each magnitude, the limits in the magnitude-redshift relations given in Table \ref{table:mag_limits}. 
Fig. \ref{fig:quality_limits} shows the magnitude limits for the $r$-band, together with the galaxies that are removed after applying the different magnitude limits.

\begin{table}
	\caption{Magnitude limits for different redshift ranges. Objects with magnitudes below these limits are not included in the training set.}
	\label{table:mag_limits}
	\centering 
	\begin{tabular}{c c}
		\hline
		\noalign{\smallskip}
		Magnitude limit &  Redshift range \\
		\noalign{\smallskip}
		\hline
		\noalign{\smallskip}
		$g$ = 11.00 + 20.00 $z_{\rm spec}$  & 0.1 < $z_{\rm spec}$ < 0.2  \\
		$g$ = 12.33 + 13.33 $z_{\rm spec}$  & 0.2 < $z_{\rm spec}$ < 0.5  \\
		$g$ = 17.75 +  2.50 $z_{\rm spec}$  & 0.5 < $z_{\rm spec}$ < 0.9  \\
		$g$ = 20.00   & $z_{\rm spec}$ > 0.9  \\
		\noalign{\smallskip}
		\hline
		\noalign{\smallskip}
		$r$ = 10.00 + 20.00 $z_{\rm spec}$  & 0.1 < $z_{\rm spec}$ < 0.2  \\
		$r$ = 12.00 + 10.00 $z_{\rm spec}$  & 0.2 < $z_{\rm spec}$ < 0.5  \\
		$r$ = 14.00 +  6.00 $z_{\rm spec}$  & 0.5 < $z_{\rm spec}$ < 1.0  \\
		$r$ = 20.00   & $z_{\rm spec}$ > 1.0 \\
		\noalign{\smallskip}
		\hline
		\noalign{\smallskip}
		$i$ = 10.00 + 20.00 $z_{\rm spec}$  & 0.1 < $z_{\rm spec}$ < 0.2  \\
		$i$ = 12.00 + 10.00 $z_{\rm spec}$  & 0.2 < $z_{\rm spec}$ < 0.4  \\
		$i$ = 14.00 +  5.00 $z_{\rm spec}$  & 0.4 < $z_{\rm spec}$ < 1.0  \\
		$i$ = 19.00   & $z_{\rm spec}$ > 1.0  \\
		\noalign{\smallskip}
		\hline
		\noalign{\smallskip}
		$z$ = 10.75 + 12.50 $z_{\rm spec}$  & 0.1 < $z_{\rm spec}$ < 0.3  \\
		$z$ = 12.25 +  7.50 $z_{\rm spec}$  & 0.3 < $z_{\rm spec}$ < 0.5  \\
		$z$ = 14.00 +  4.00 $z_{\rm spec}$  & 0.5 < $z_{\rm spec}$ < 1.0  \\
		$z$ = 18.00   & $z_{\rm spec}$ > 1.0  \\
		\noalign{\smallskip}
		\hline
		\noalign{\smallskip}
		$y$ =  9.25 + 17.50 $z_{\rm spec}$   & 0.1 < $z_{\rm spec}$ < 0.3  \\
		$y$ = 12.25 +  7.50 $z_{\rm spec}$   & 0.3 < $z_{\rm spec}$ < 0.5  \\
		$y$ = 14.00 +  4.00 $z_{\rm spec}$   & 0.5 < $z_{\rm spec}$ < 1.0  \\
		$y$ = 18.00    & $z_{\rm spec}$ > 1.0  \\
		\noalign{\smallskip}
		\hline
	\end{tabular}
\end{table}

Figure \ref{fig:distance_distribution} shows the distribution of distances between the PS1 and SDSS objects in the cleaned catalogue. The scale is logarithmic to highlight that there is a small tail of objects far away from the SDSS position, as allowed from our large search radius (30\arcsec).  
Since our goal is to keep only the objects for which the match is secure, we removed from our sample all the objects whose distances to the SDSS positions were larger than 1\arcsec. 
These were 2368 objects out of 2\,316\,092 (corresponding to $\sim0.10\%$), resulting in a training set $\mathcal{T}$ with 2\,313\,724 objects. This conservative approach allows us to safely use the SDSS spectroscopic redshift with the PS1 magnitudes.

\subsection{Final training set}\label{ssec:summary_training}

The final training set $\mathcal{T}$ contains 2\,313\,724 galaxies. For each galaxy, the training set provides the spectroscopic redshift $z_{\rm spec}$ obtained from the catalogue of \citet{Beck2016}, and the 5 features ($g-r$, $r-i$, $i-z$, $z-y$, and $r$) obtained as explained in Sect. \ref{ssec:feature_computation}. The features that could not be calculated due to a missing magnitude were set to a default value (-999).

\begin{figure}[t]
	\centering
	\includegraphics[width=0.99\columnwidth]{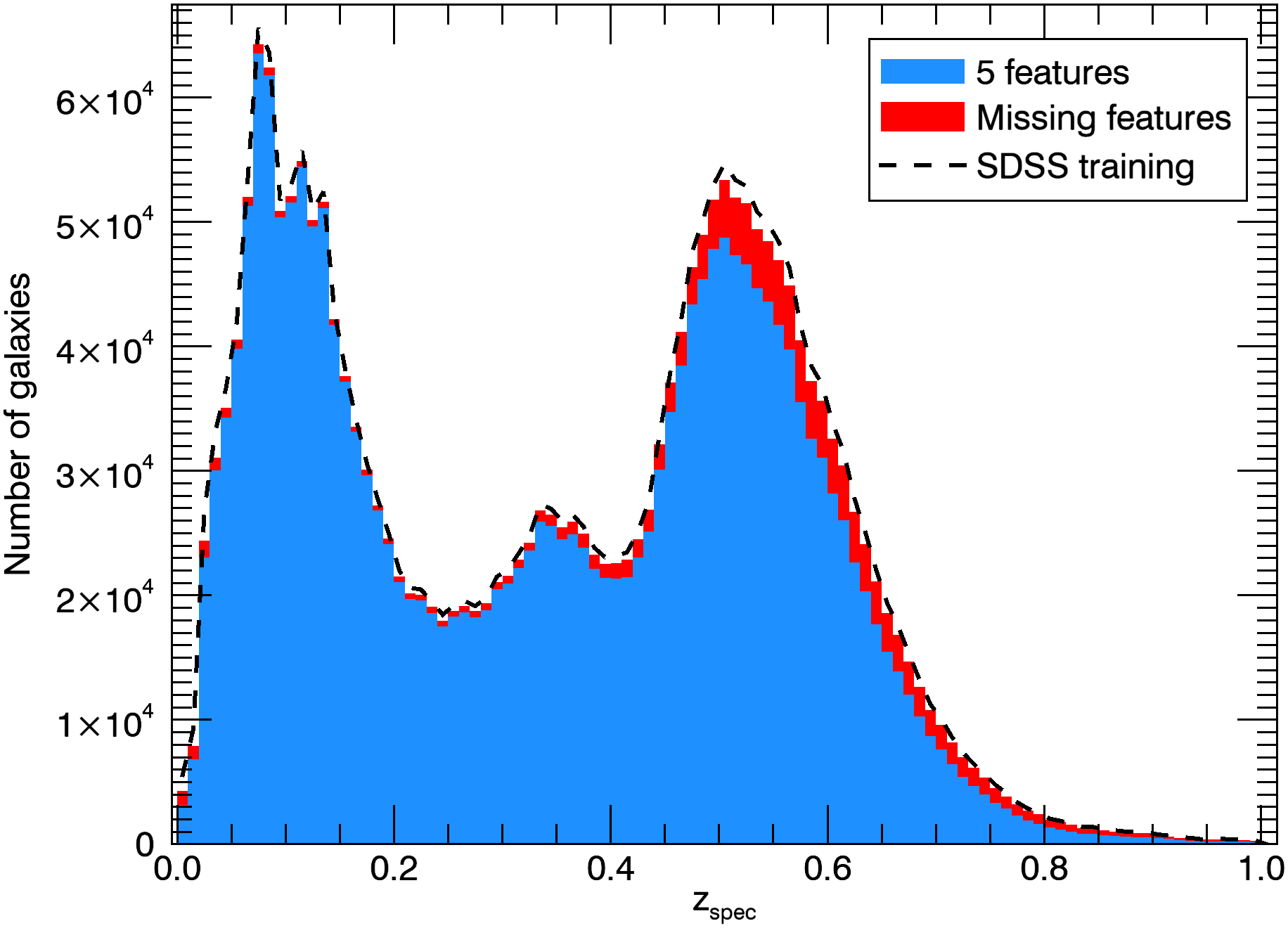}
	\caption{Distribution of the  spectroscopic redshifts in our training set. In blue we show the galaxies that have the 5 features available ($\mathcal{T}_5$). In red, the remaining galaxies. The dashed black line shows the redshift distribution of the original SDSS training set.
	}
	\label{fig:zspec_distribution}
\end{figure}

The redshift distribution of the galaxies in $\mathcal{T}$ is shown in Fig. \ref{fig:zspec_distribution}, where $\mathcal{T}$ has been divided into two subsets: the galaxies that have the 5 features available ($\mathcal{T}_5$, in blue), and the galaxies that have one or more features missing (in red).  
For comparison, Fig.~\ref{fig:zspec_distribution} also shows the spectroscopic redshift distribution of the original SDSS training set, which contained 2\,379\,096 galaxies.

The PS1 training set that we are presenting in this Section is smaller than the original SDSS one, as expected when doing a match between different catalogues. This difference can be due to errors in the astrometry or photometry of the two surveys, as well as to intrinsic limits of our match methodology. However, we stress that we tried to use a conservative approach in which the number of galaxies is smaller, but the robustness of the match is favoured. This result was reached while loosing less than 3\% of the original galaxies.

\begin{figure*}
	\centering
	\subfigure{\includegraphics[width=0.66\columnwidth]{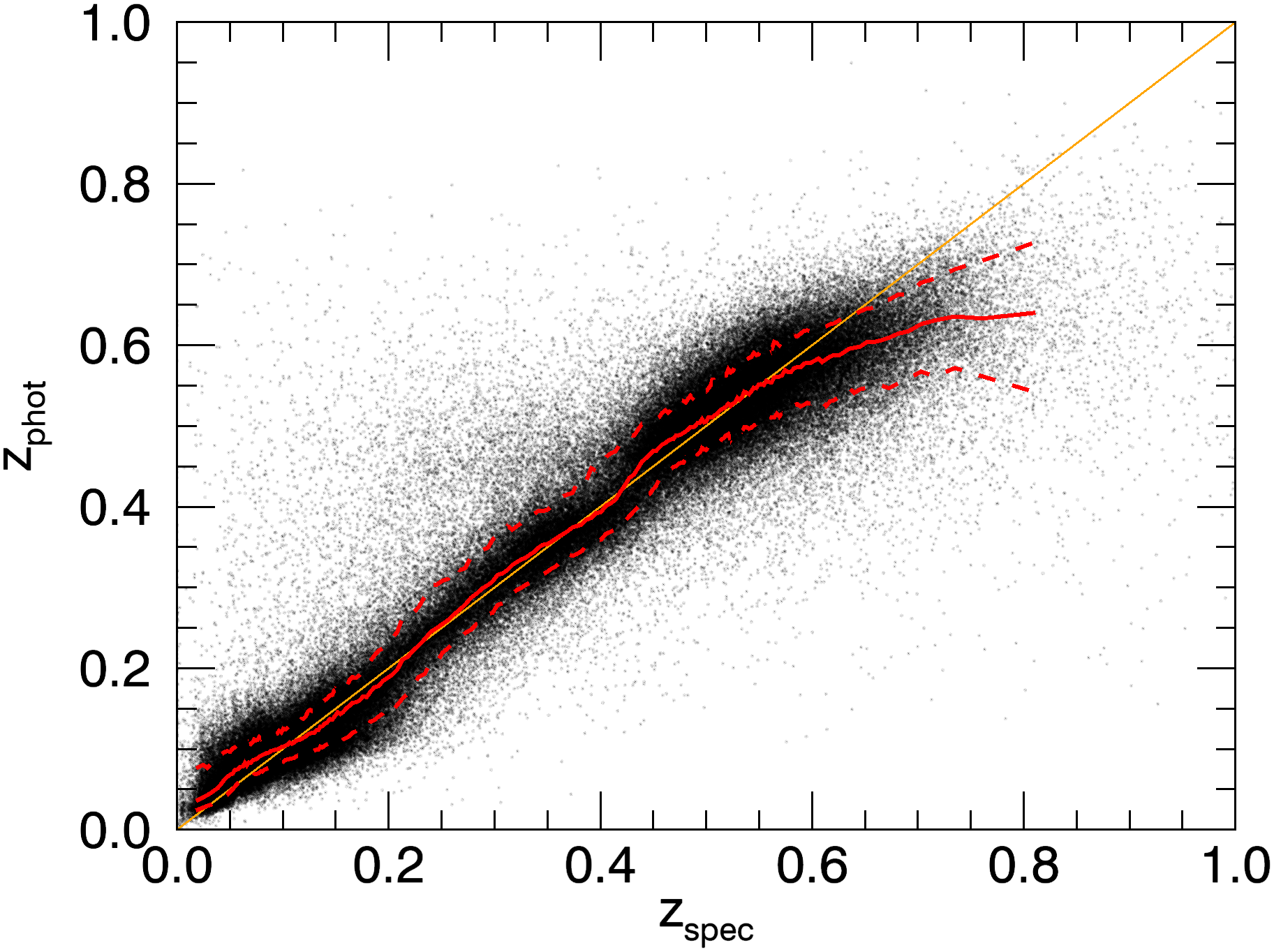}}
	\subfigure{\includegraphics[width=0.66\columnwidth]{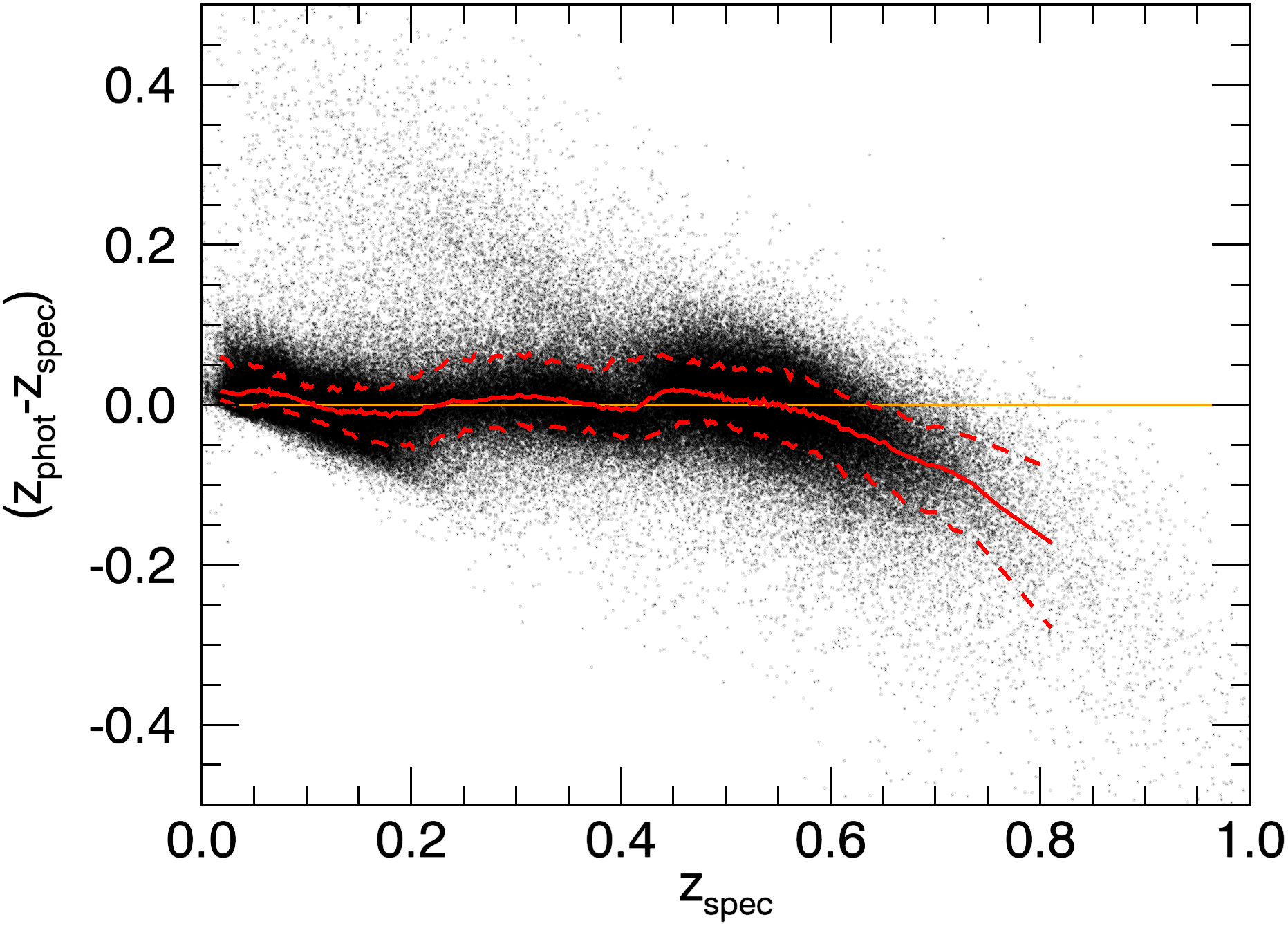}}
	\subfigure{\includegraphics[width=0.66\columnwidth]{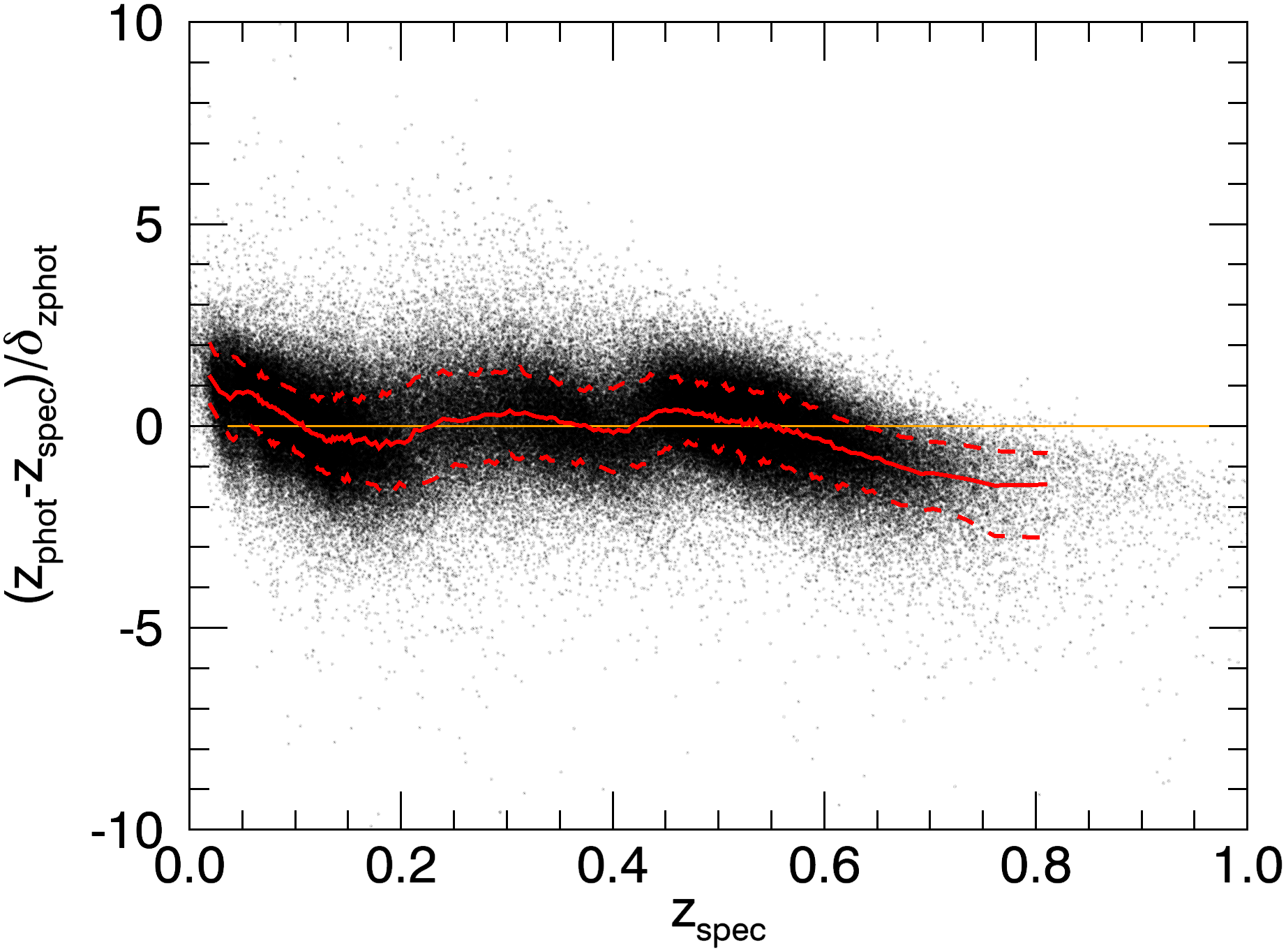}}
	\caption{Photometric redshift $z_{\rm phot}$ (left),  redshift error $z_{\rm phot}-z_{\rm spec}$ (middle), and error divided by the estimation of the error provided by the method $(z_{\rm phot}-z_{\rm spec})/\delta_{z_{\rm phot}}$ (right) as a function of the spectroscopic redshift $z_{\rm spec}$. The black dots represent each individual galaxy of the 10 different $\mathcal{T}_{\rm test}$ sets. Only the galaxies with the 5 magnitudes available were included. The red solid and dotted lines represent the median and the 68\% confidence regions, respectively, computed for groups of 1000 galaxies with consecutive $z_{\rm spec}$. The orange line shows $z_{\rm phot}=z_{\rm spec}$.}
	\label{fig:zphot_zspec}
\end{figure*}

\section{Performance of the method}\label{sec:results}

\subsection{Overall redshift precision}\label{ssec:general_results}
To evaluate the performance of the proposed method, we randomly divided the training set $\mathcal{T}$ into two disjoint subsets of different sizes: $\mathcal{T}_{\rm train}$ and $\mathcal{T}_{\rm test}$. 
The former is used as training set for estimating the photometric redshift of the galaxies in the latter. 
We chose the sizes to be 99\% and 1\% of $\mathcal{T}$, respectively. We repeated the experiment 10 times to analyse the stability of the results. Figure \ref{fig:zphot_zspec} shows the photometric redshift $z_{\rm phot}$, the actual error $z_{\rm phot}-z_{\rm spec}$, and the error divided by the estimation of the error provided by the method $(z_{\rm phot}-z_{\rm spec})/\delta_{z_{\rm phot}}$ as a function of the spectroscopic redshift for the galaxies in the 10 different $\mathcal{T}_{\rm test}$ sets having the 5 magnitudes available. The photometric redshift follows quite well the spectroscopic redshift, especially in the intermediate redshift range ($0.1<z_{\rm spec}<0.6$), where the average bias $\Delta z = |z_{\rm phot}-z_{\rm spec}|$ is below 0.02 or 0.5$\delta_{z_{\rm phot}}$. 

For high-redshift galaxies ($z_{\rm spec}>0.6$), the method tends to underestimate the redshift, and also presents a higher scatter. This behaviour was also observed in \citet{Beck2016}. The increased scatter is due to the low number of high-redshift galaxies in the training set. The negative bias is an Eddington bias produced by the limited depth of the PS1 survey: close to the detection  limit, over-luminous galaxies are preferentially detected, yielding a bias towards lower redshifts.
In this redshift range 54\% of the galaxies are within $\pm\delta_{z_{\rm phot}}$ and 86\% are within $\pm2\delta_{z_{\rm phot}}$. The percentage of galaxies in this range whose redshift is estimated via extrapolation is 0.10\%, six times higher than the value in the intermediate redshift range (0.017\%). For low-redshift galaxies ($z_{\rm spec}<0.1$), the method tends to overestimate the redshift, as in \citet{Beck2016}. In this redshift range 65\% of the galaxies are within $\pm\delta_{z_{\rm phot}}$ and 94\% are within $\pm2\delta_{z_{\rm phot}}$. The percentage of galaxies in this range whose redshift is estimated via extrapolation is 0.074\%, higher than in the intermediate redshift range, but lower than in the high redshift range. 

In order to quantitatively compare the average performance of the proposed method to the one obtained in \citet{Beck2016}, we use the same definition of the normalized redshift estimation error, that is $\Delta z_{\rm norm} = \frac{z_{\rm phot}-z_{\rm spec}}{1+z_{\rm spec}}$. After iteratively removing the outliers, defined as $|\Delta z_{\rm norm}|>3\sigma(\Delta z_{\rm norm})$, the average bias of our method is 
$\overline{\Delta z_{\rm norm}}=-2.01 \times 10^{-4}$, the standard deviation is $\sigma(\Delta z_{\rm norm})=0.0298$, and the outlier rate is $P_o=4.32\%$, 
when calculated on the ensemble of results from the 10 experiments. The differences between the 10 experiments were negligible. For reference, the results reported by \citet{Beck2016} were $\overline{\Delta z_{\rm norm}}=5.84 \times 10^{-5}$, $\sigma(\Delta z_{\rm norm})=0.0205$, and $P_o=4.11\%$, which are of the same order, but slightly better than ours. We note however that they were calculated using a different training set, so they are not directly comparable. A direct comparison is presented in Sect. \ref{ssec:SDSS}.

\begin{figure}[]
	\centering
	\includegraphics[width=0.99\columnwidth]{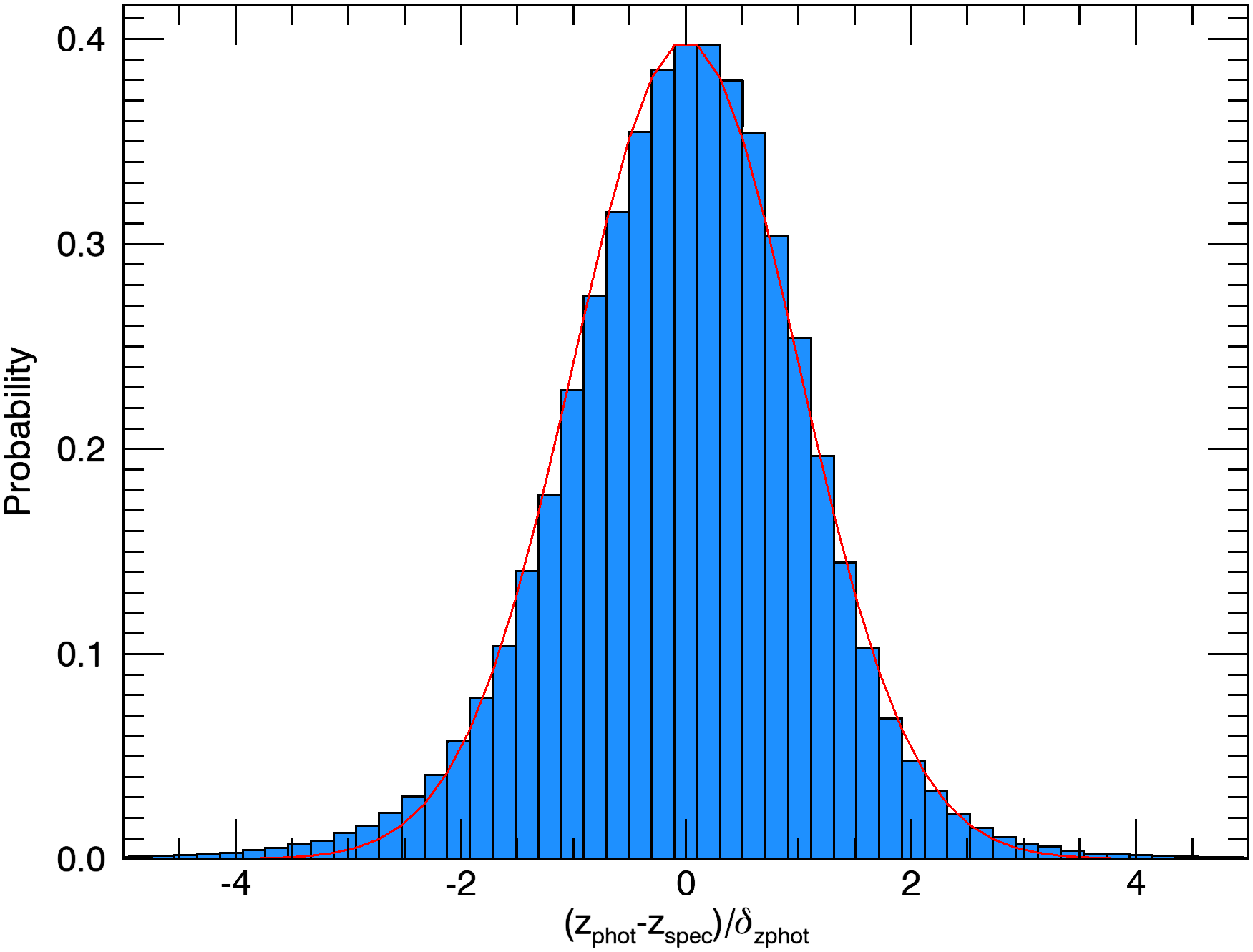}
	\caption{Normalized histogram of $z_{\rm phot}-z_{\rm spec}/\delta_{z_{\rm phot}}$. For reference, the red line shows a standard Gaussian distribution.}
	\label{fig:histogram}
\end{figure}

Figure \ref{fig:histogram} shows the normalized histogram of $(z_{\rm phot}-z_{\rm spec})/\delta_{z_{\rm phot}}$ together with a standard normal distribution. The two distributions are well in agreement, apart from a small bias \citep[as in][cfr. their fig. 4]{Beck2016}. This indicates that the estimated errors $\delta_{z_{\rm phot}}$ represent quite well the accuracy of the redshift estimation.

\subsection{Impact of the photometric errors}\label{ssec:photometric_errors}
The errors in the measurements of the photometric magnitudes have an impact in the final accuracy of the estimated redshift. To evaluate this effect, we classified the galaxies into five different classes according to their photometric errors. Class 1 includes galaxies with low photometric errors, and classes 2-5 include galaxies with progressively higher errors. The error limits for the different classes were manually chosen and are given in Table \ref{table:error_limits}.  
We also define an additional class E, which includes the galaxies whose redshift is estimated via an extrapolation of Eq. \ref{eq:linear_model}. This occurs when the galaxy features lie outside the bounding box of its nearest neighbours, as mentioned in Sect. \ref{ssec:linear_regression}. 

The photometric error corresponding to the $r$-band Kron magnitude ($\Delta r$) is directly obtained from the query to the PS1 database ({\tt rKronMagErr} in Table \ref{table:query_output}). The photometric errors in the four aperture colours are obtained from the errors in the corresponding aperture fluxes. If $f_g$ is the aperture flux in the $g$ band and $\Delta f_g$ the corresponding error, which are obtained from the query ({\tt rgc6flxR} and {\tt rgc6flxErrR} in Table \ref{table:query_output}), the error in the $g$-band aperture magnitude is calculated as $\Delta g = 2.5 \log(e) \times \Delta f_g/f_g$, and analogously for the other aperture magnitudes. The error in the aperture colours is thus calculated as $\Delta(g-r) = \sqrt{(\Delta g)^2+(\Delta r)^2}$, and similarly for the other colours.

Table \ref{table:errors} summarizes the performance of the redshift estimation for the different photometric classes. 
The bias is very close to 0 for all the classes, being positive for class 1 and increasingly negative for classes 2 to 5. This results in a slightly negative bias for the whole sample ($\overline{\Delta z_{\rm norm}}=-2.01 \times 10^{-4}$) since, even though the number of class 1 galaxies dominates, the negative bias of the other classes is higher in absolute value. The standard deviation of the normalized redshift estimation error $\sigma(\Delta z_{\rm norm})$ increases for higher photometric errors, as expected, and the same occurs with the outlier rate. For the galaxies in class E, the bias and  $\sigma(\Delta z_{\rm norm})$ is higher than for the other classes.

The defined photometric classes can be used to filter out, if needed, the galaxies for which the redshift estimation is less precise.

\begin{table}
	\caption{Photometric error limits for the defined classes. A galaxy belongs to a given class if its five photometric errors are below the specified values for that class and it is the lowest possible class. Class 5 contains galaxies for which one or more of the photometric errors are above the limits corresponding to class 4.}
	\label{table:error_limits}
	\centering 
	\small
	\begin{tabular}{c | c c c c c}
		\hline
		\noalign{\smallskip}
		Class &  $\Delta r_{\rm max}$  & $\Delta(g-r)_{\rm max}$ & $\Delta(r-i)_{\rm max}$ & $\Delta(i-z)_{\rm max}$ & $\Delta(z-y)_{\rm max}$\\
		\noalign{\smallskip}
		\hline
		\noalign{\smallskip}
		1 &    0.05 &    0.10 &    0.05 &    0.05 &    0.10 \\ 
		2 &    0.10 &    0.20 &    0.10 &    0.10 &    0.20 \\ 
		3 &    0.15 &    0.30 &    0.15 &    0.15 &    0.30 \\ 
		4 &    0.25 &    0.50 &    0.25 &    0.25 &    0.50 \\ 
		\noalign{\smallskip}
		\hline
	\end{tabular}
\end{table}

\begin{table}
	\caption{Average normalized redshift estimation bias $\overline{\Delta z_{\rm norm}}$, standard deviation $\sigma(\Delta z_{\rm norm})$ and outlier rate $P_o$ for the defined photometric classes. These quantities were calculated after iteratively removing the outliers, defined as $|\Delta z_{\rm norm}|>3\sigma(\Delta z_{\rm norm})$. The number of galaxies $N$, from the 10 different $\mathcal{T}_{\rm test}$ sets, belonging to each class is also indicated. }
	\label{table:errors}
	\centering 
	\begin{tabular}{c | c c c c}
		\hline
		\noalign{\smallskip}
		Class &  $\overline{\Delta z_{\rm norm}}$ &  $\sigma(\Delta z_{\rm norm})$ &    $P_o$ & $N$ \\
		\noalign{\smallskip}
		\hline
		\noalign{\smallskip} 

1 &            1.31 $\times 10^{-4}$ &  0.0265 &    2.90 &     126374 \\ 
2 &           -6.29 $\times 10^{-4}$ &  0.0329 &    3.84 &      46458 \\ 
3 &           -8.53 $\times 10^{-4}$ &  0.0346 &    5.01 &      18648 \\ 
4 &           -2.27 $\times 10^{-3}$ &  0.0399 &    7.82 &      12641 \\ 
5 &           -2.34 $\times 10^{-3}$ &  0.0448 &    8.99 &      11953 \\ 
\noalign{\smallskip}
\hline
\noalign{\smallskip} 
E &            8.36 $\times 10^{-3}$ &  0.1095 &    7.89 &         76 \\
\noalign{\smallskip}
\hline
\noalign{\smallskip} 
all &        -2.01 $\times 10^{-4}$ &  0.0298 &    4.32 &     216150 \\
\noalign{\smallskip}
\hline
	\end{tabular}
\end{table}

\begin{figure*}
	\centering
	\subfigure{\includegraphics[width=0.66\columnwidth]{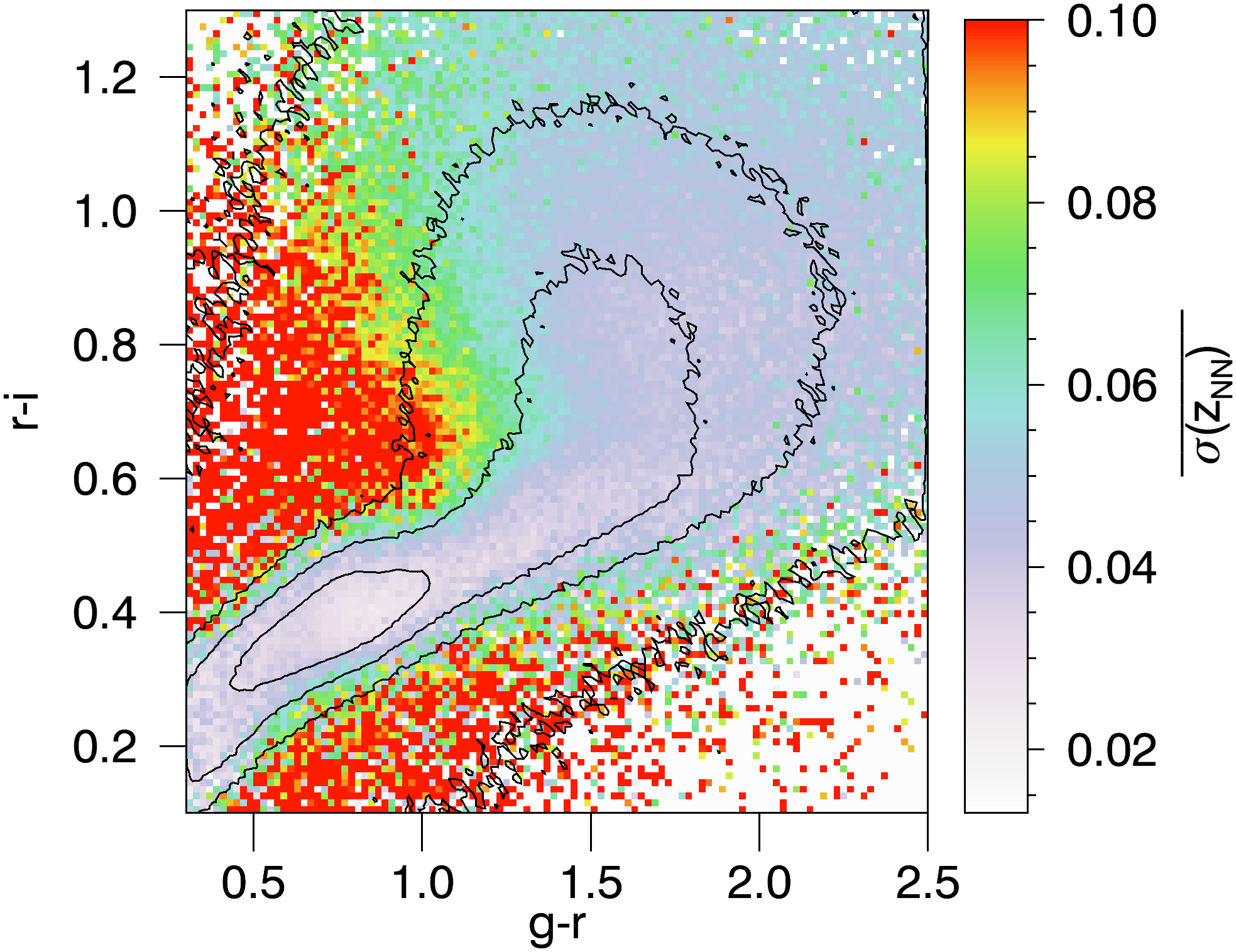}}
	\subfigure{\includegraphics[width=0.66\columnwidth]{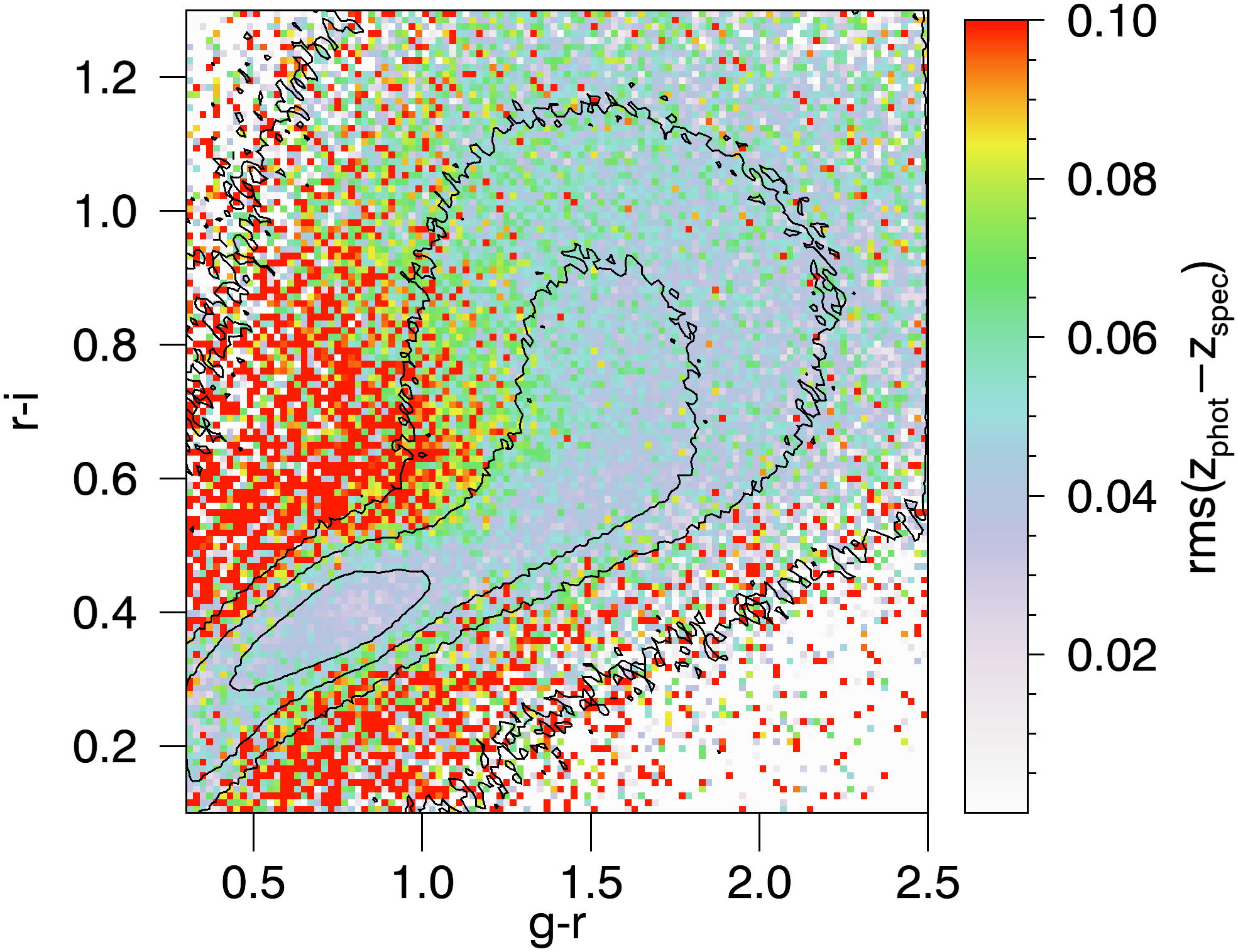}}
	\subfigure{\includegraphics[width=0.66\columnwidth]{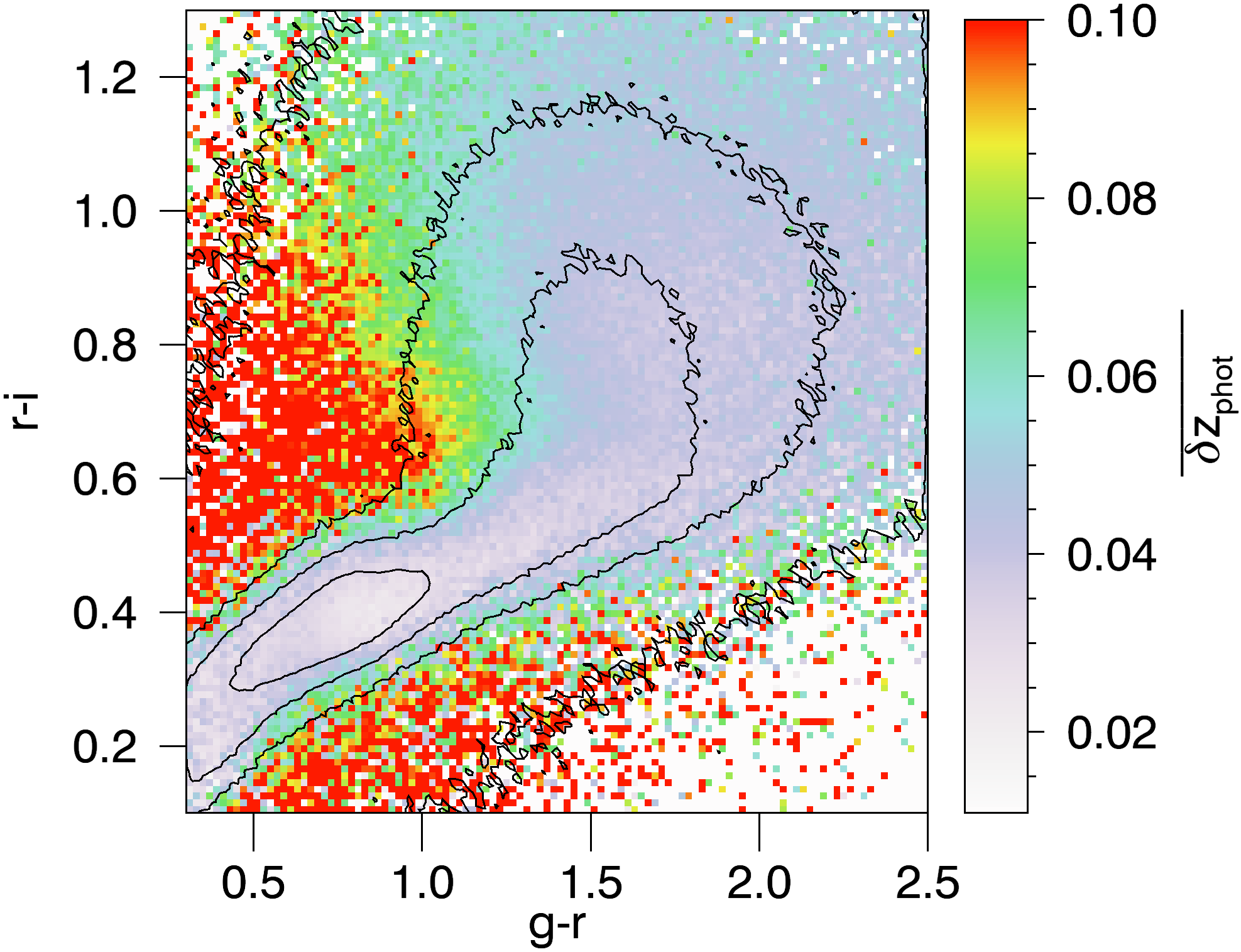}}
	\caption{Photometric redshift results in the 10 different $\mathcal{T}_{\rm test}$ sets as a function of the $r-i$ and the $g-r$ colours. The left panel shows the average standard deviation of the redshifts of the nearest neighbours $\sigma(z_{\rm NN})$, the middle panel shows the rms of the actual error $z_{\rm phot}-z_{\rm spec}$, and the right panel shows the average estimated errors $\delta_{z_{\rm phot}}$. For easier comparison, the scale in the three panels was set between 0 and 0.1, with the red colour indicating errors that are bigger or equal than 0.1. For reference, the black lines represent the contours of the galaxy count distribution of the training set $\mathcal{T}_{5}$, with the four displayed contours corresponding to 1000, 300, 100 and 10 galaxies per colour bin.}
	\label{fig:error_map}
\end{figure*}

\begin{figure*}
	\centering
	\subfigure{\includegraphics[width=0.66\columnwidth]{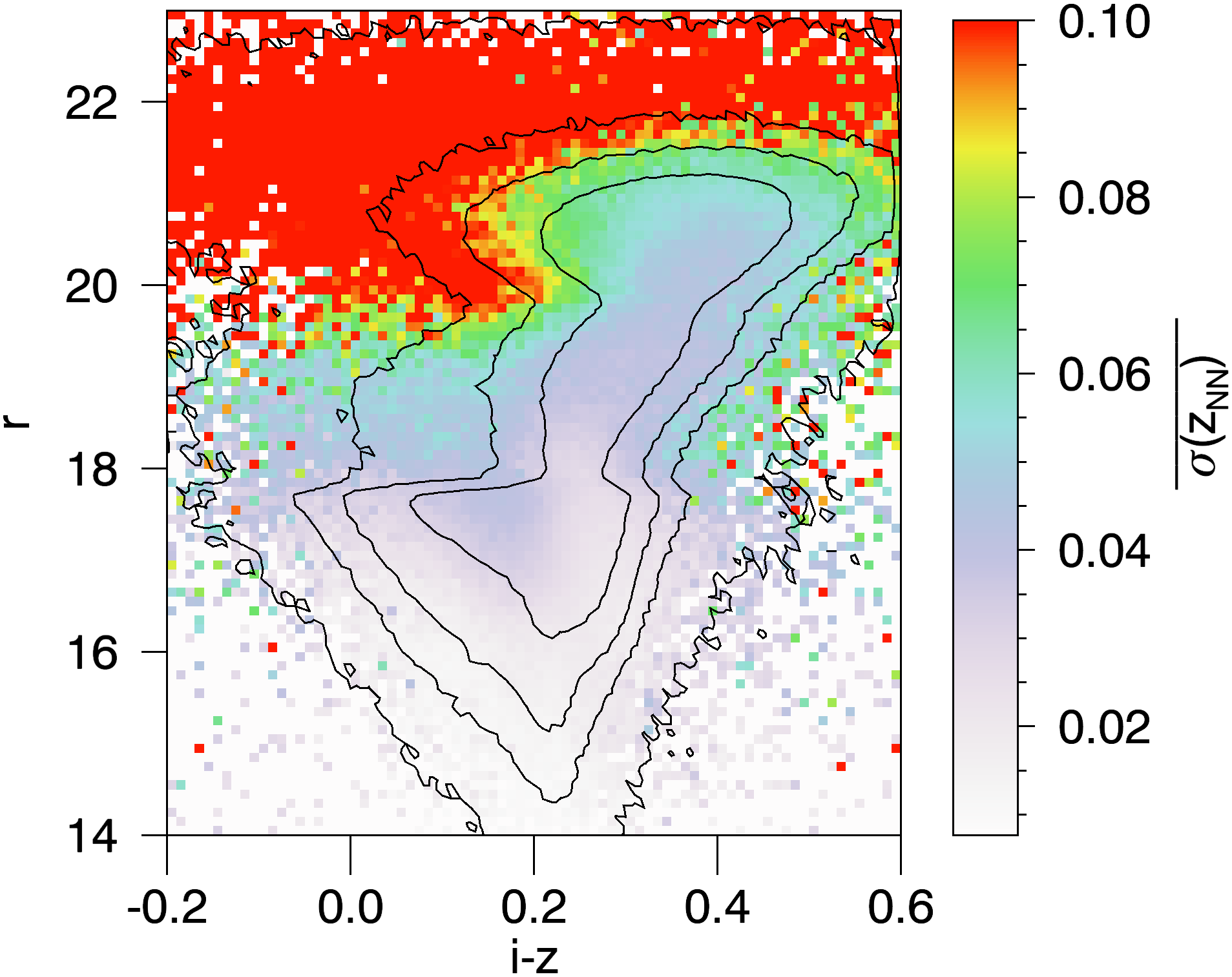}}
	\subfigure{\includegraphics[width=0.66\columnwidth]{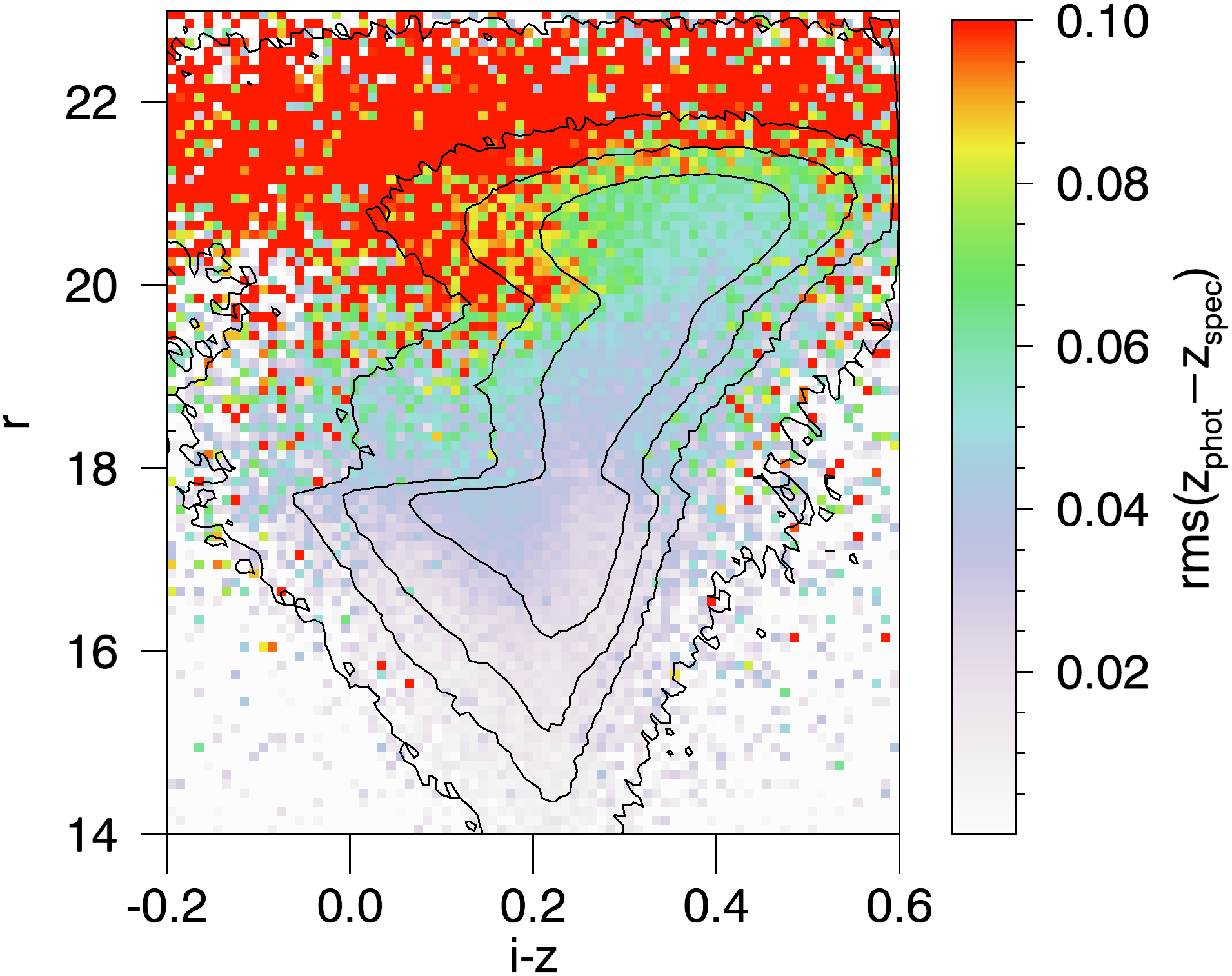}}
	\subfigure{\includegraphics[width=0.66\columnwidth]{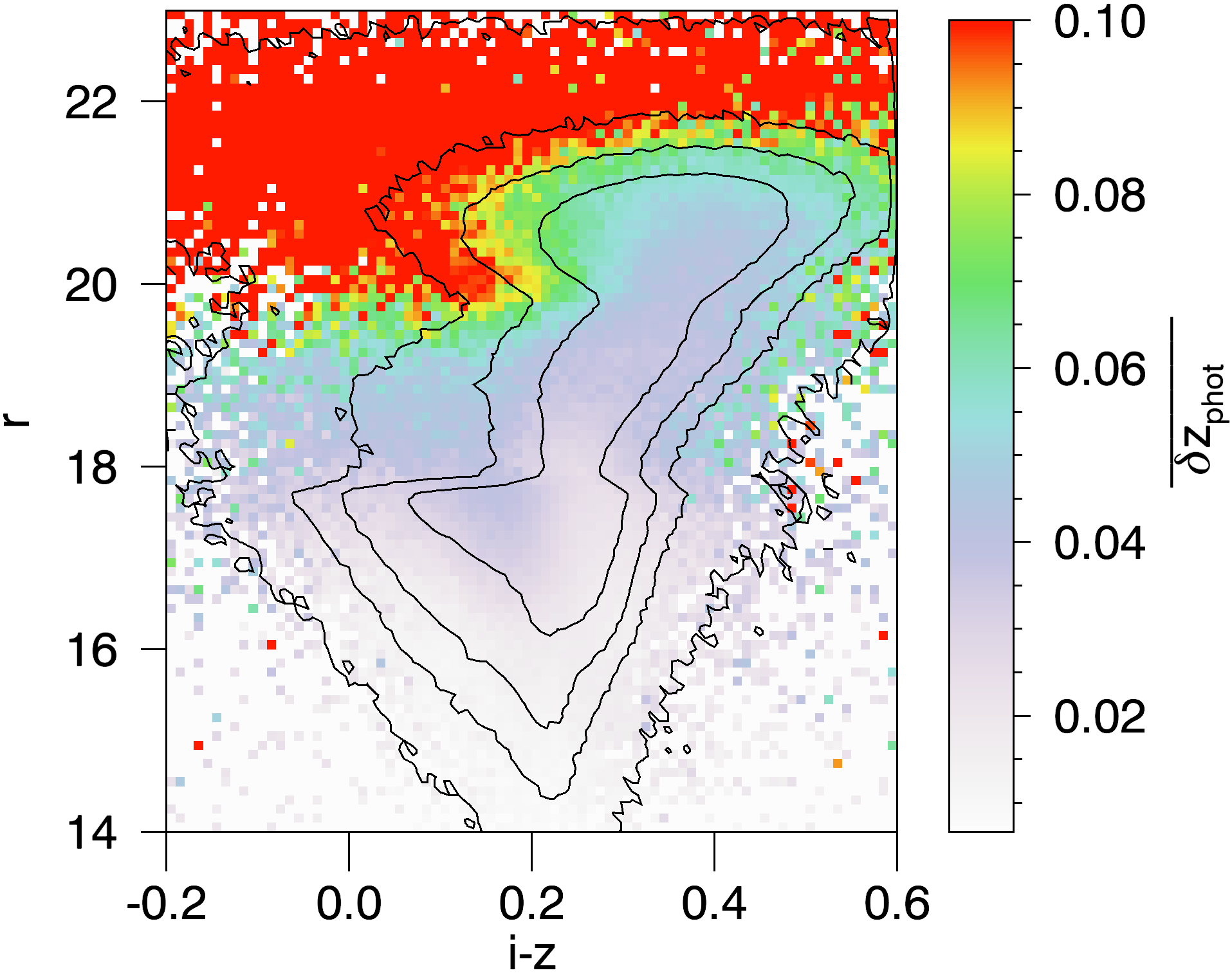}}
	\caption{Photometric redshift results in the 10 different $\mathcal{T}_{\rm test}$ sets as a function of the $r$ magnitude and the $i-z$ colour. The left panel shows the average standard deviation of the redshifts of the nearest neighbours $\sigma(z_{\rm NN})$, the middle panel shows the rms of the actual error $z_{\rm phot}-z_{\rm spec}$, and the right panel shows the average estimated errors $\delta_{z_{\rm phot}}$. The colour scale and the black contours are set as in Fig. \ref{fig:error_map}.}
	\label{fig:error_map_r_iz}
\end{figure*}

\subsection{Impact of the position in the colour-magnitude space}\label{ssec:error_maps}
The position of the galaxy in the $D$-dimensional feature space also has an effect on the redshift estimation error. Galaxies situated in dense regions are expected to have smaller errors, since their neighbours will be very close to them in the $D$-dimensional colour-magnitude space, and will probably have similar redshifts. On the contrary, galaxies situated in sparse regions will have larger errors in the redshift estimation, since their neighbours will be further away in the colour-magnitude space, and will probably have a bigger dispersion in their redshifts.

To characterize this effect we have computed several error maps that provide the redshift estimation errors as a function of the position in the $D$-dimensional colour-magnitude space. These error maps can be used to filter out, if required, the regions in the colour-magnitude space that have larger errors.

\begin{figure*}[]
	\centering
	\includegraphics[width=1.00\columnwidth]{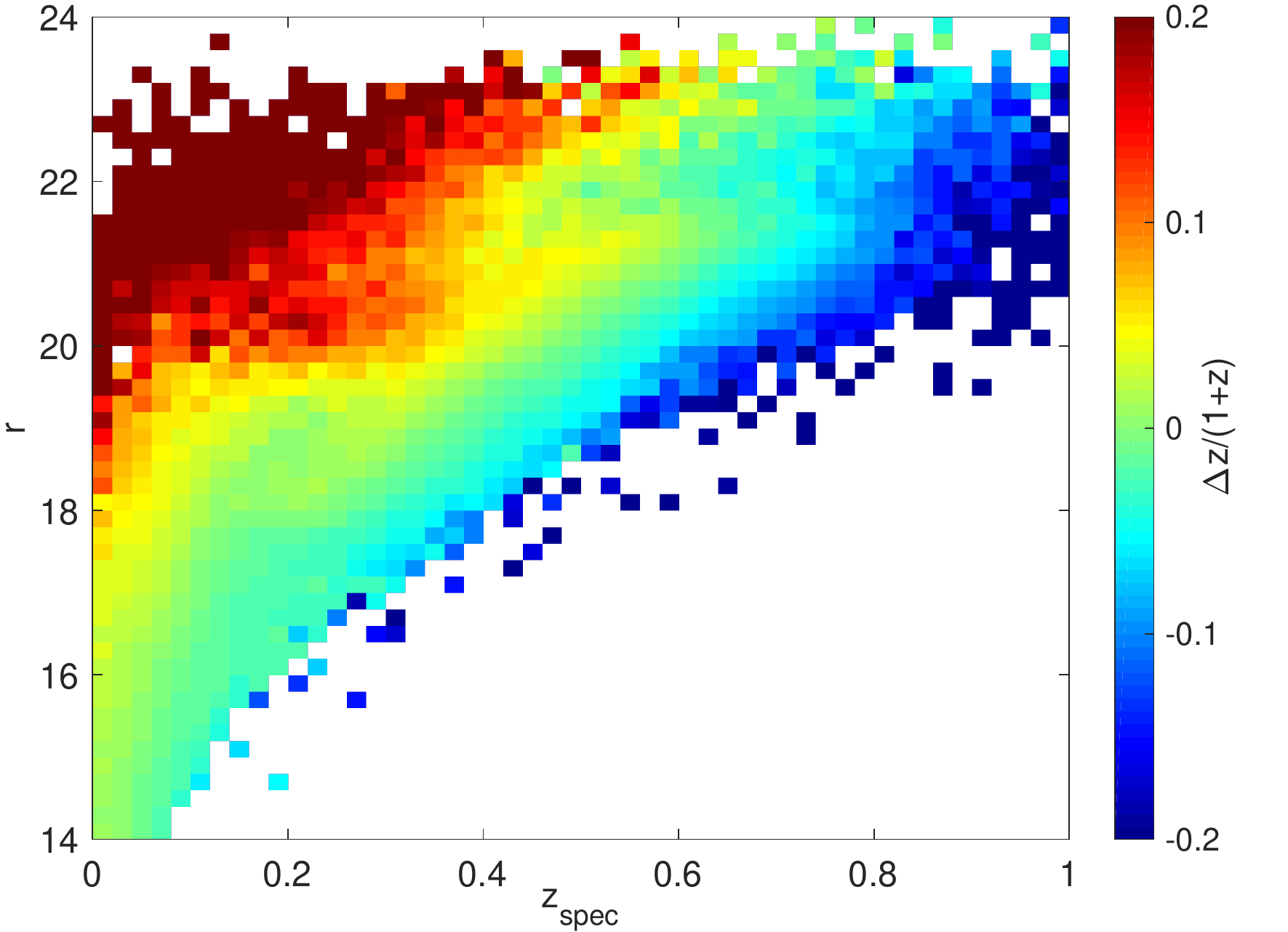}\label{fig:bias_vs_zspec_and_r}
	\includegraphics[width=0.99\columnwidth]{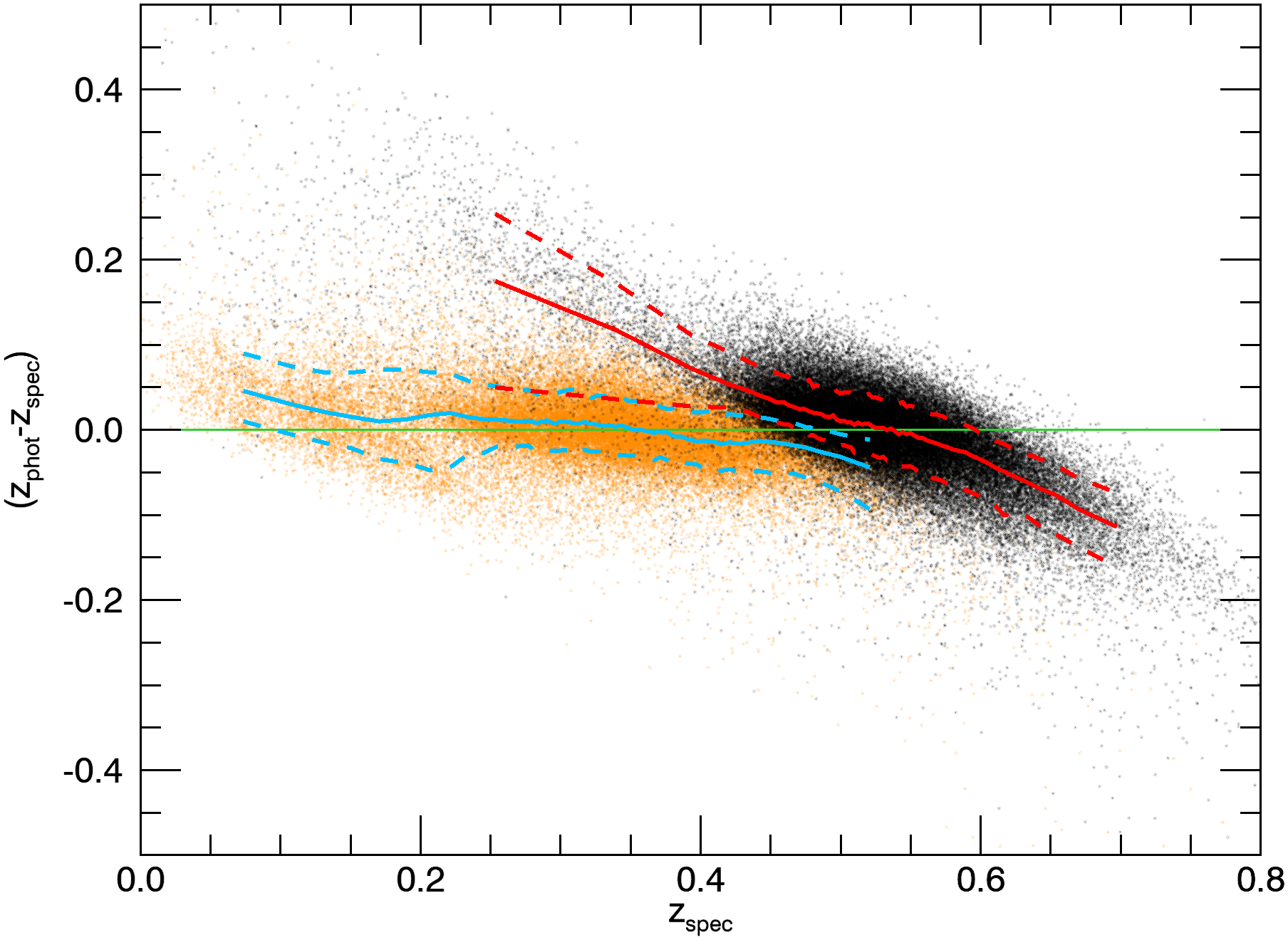}\label{fig:error_18_20_and_20_21}
	\caption{Left panel: Average normalized error $\Delta z_{\rm norm} = (z_{\rm phot}-z_{\rm spec})/(1+z_{\rm spec})$ in the 10 different $\mathcal{T}_{\rm test}$ sets as a function of the magnitude $r$ and the spectroscopic redshift $z_{\rm spec}$. Right panel: redshift error $z_{\rm phot}-z_{\rm spec}$ as a function of the spectroscopic redshift $z_{\rm spec}$ for galaxies with $18<r<20$ (orange dots) and $20<r<21$ (black dots). Each dot represents an individual galaxy of the 10 different $\mathcal{T}_{\rm test}$ sets. The thick solid and dotted lines represent the median and the 68\% confidence regions, respectively, computed in small $z_{\rm spec}$ intervals for the galaxies with $18<r<20$ (blue lines) and $20<r<21$ (red lines). The green line shows $z_{\rm phot}=z_{\rm spec}$.}
	\label{fig:effect_of_r}
\end{figure*}

Figure \ref{fig:error_map} illustrates this effect in the $g-r$ and $r-i$ colour plane. 
The colour maps in this figure show three measurements of the redshift estimation error as a function of $g-r$ and $r-i$: the average standard deviation of the redshifts of the nearest neighbours $\sigma(z_{\rm NN})$, the root mean square (rms) of the actual error $z_{\rm phot}-z_{\rm spec}$, and the average estimated errors $\delta_{z_{\rm phot}}$. The different error measurements show a similar behaviour. The estimated error $\delta_{z_{\rm phot}}$ is closely related to the actual error, which further supports that it is a good estimator of the error, as previously shown in Fig. \ref{fig:histogram}. As expected, both errors are clearly correlated with the deviation of the redshifts of the nearest neighbours. In the regions where the dispersion is higher, the redshift estimation has a bigger error. 
The contour lines in Fig. \ref{fig:error_map} represent the galaxy count distribution of the training set $\mathcal{T}_{5}$. By comparing these contours with the background error maps, 
we see that there is a clear correlation between the photometric redshift errors and the galaxy count distribution: denser regions yield smaller errors and sparser regions yield bigger errors.

Figure \ref{fig:error_map_r_iz} shows the same three measurements of the redshift estimation error shown in Fig. \ref{fig:error_map} as a function of $r$ and $i-z$. The behaviour is similar to that observed in Fig. \ref{fig:error_map}, with an estimated error $\delta_{z_{\rm phot}}$ closely following the deviation of the redshifts of the nearest neighbours $\sigma(z_{\rm NN})$ and the actual error $z_{\rm phot}-z_{\rm spec}$. The contour lines in this figure show the galaxy count distribution as a function of $r$ and $i-z$. By comparing these contours with the error maps we see again a correlation between the two, although less clear than in Fig. \ref{fig:error_map}. This is due to an additional effect that is analysed next: fainter galaxies tend to have larger errors than brighter galaxies. 

In fact, one of the input features of the proposed method depends directly on the $r$-band magnitude of the galaxies, and thus, on their apparent brightness. This feature has a strong impact on the estimation of the photometric redshift, as shown in Fig. \ref{fig:effect_of_r}.
The left panel of Fig. \ref{fig:effect_of_r} shows the average normalized error $\Delta z_{\rm norm} = (z_{\rm phot}-z_{\rm spec})/(1+z_{\rm spec})$ as a function of the magnitude $r$ and the spectroscopic redshift $z_{\rm spec}$. While for brighter galaxies ($r<20$) the average normalized error is below 0.1, for fainter galaxies ($r>20$) the error increases significantly, especially for galaxies at $z<0.4$ or $z>0.8$.
The right panel of Fig. \ref{fig:effect_of_r} shows the redshift error $z_{\rm phot}-z_{\rm spec}$ for bright ($18<r<20$) and faint ($20<r<21$) galaxies. While the redshift of brighter galaxies is well estimated, with a small bias both at low and high redshift, the error for fainter galaxies is higher, especially when the true redshift is far from $~0.5-0.6$.

\subsection{Impact of missing features}\label{ssec:missing_feature}
The proposed method is also able to work when one or several features are missing. When this occurs, the performance of the method degrades, with an increased scatter in the photometric versus spectroscopic redshift relation. Missing features usually appear in very faint galaxies, but can also occur in brighter galaxies due to photometric measurement errors. In our training set $\mathcal{T}$ with 2\,313\,724 galaxies, the $r$ Kron magnitude is missing from 48\,416 galaxies (2.1\%), and the $g-r$, $r-i$, $i-z$ and $z-y$ aperture colours are missing from 144\,352 (6.2\%), 51\,150 (2.2\%), 52\,357 (2.3\%), and 57\,350 (2.5\%) galaxies, respectively. Most of these galaxies are faint. For example, approximately 91\% of the galaxies without the $g-r$ aperture colour have an $r$ Kron magnitude $r > 20$, while only 9\% are brighter than $r = 20$. 

To evaluate the effect of a missing feature independently of the position of the galaxy in the magnitude-colour space, we artificially removed one of the features from our training set $\mathcal{T}_5$, and repeated the experiment described in Section \ref{ssec:general_results}, using the four remaining features for both the test and training subsets. The results are similar to those presented in Fig. \ref{fig:zphot_zspec}, but with a higher scatter. Table \ref{table:errors_misssing_features} summarizes the results obtained when the different features are removed. When the $r$ Kron magnitude is removed, the standard deviation of the normalized bias is $\sigma(\Delta z_{\rm norm})=0.0365$, 22\% higher than when using the 5 features ($\sigma(\Delta z_{\rm norm})=0.0298$). The effect is smaller when one of the aperture colours is removed, with an increase of 13\%, 11\%, 4\%, and 1\% in $\sigma(\Delta z_{\rm norm})$ for $g-r$, $r-i$, $i-z$, and $z-y$, respectively. This indicates that, among the five features, the $r$ Kron magnitude has the strongest effect in the determination of the photometric redshifts, while the aperture colours play a weaker role. On the other hand, the average bias remains small, and the outlier rate is not much affected by the removal of any of the colour features, but increases when the $r$ magnitude is removed.

\begin{table}
	\caption{Average normalized redshift estimation bias $\overline{\Delta z_{\rm norm}}$, standard deviation $\sigma(\Delta z_{\rm norm})$ and outlier rate $P_o$ for the experiments in which one of the features is removed. These quantities were calculated after iteratively removing the outliers, defined as $|\Delta z_{\rm norm}|>3\sigma(\Delta z_{\rm norm})$. }
	\label{table:errors_misssing_features}
	\centering 
	\begin{tabular}{c | c c c }
		\hline
		\noalign{\smallskip}
		Removed feature &  $\overline{\Delta z_{\rm norm}}$ &  $\sigma(\Delta z_{\rm norm})$ &    $P_o$  \\
		\noalign{\smallskip}
		\hline
		\noalign{\smallskip}
		 $r$ 	& 		1.04 $\times 10^{-3}$ &  0.0365 &    5.64  \\ 
		 $g-r$  &        -2.14 $\times 10^{-4}$ &  0.0338 &    4.12   \\
		 $r-i$  &        -1.09 $\times 10^{-3}$ &  0.0330 &    4.36    \\ 
		 $i-z$  &        -1.58 $\times 10^{-4}$ &  0.0310 &    4.13   \\ 
		 $z-y$  &        -2.74 $\times 10^{-4}$ &  0.0302 &    4.21  \\ 

		\noalign{\smallskip}
		\hline
	\end{tabular}
\end{table}

\begin{figure*}
	\centering
	\subfigure{\includegraphics[width=0.66\columnwidth]{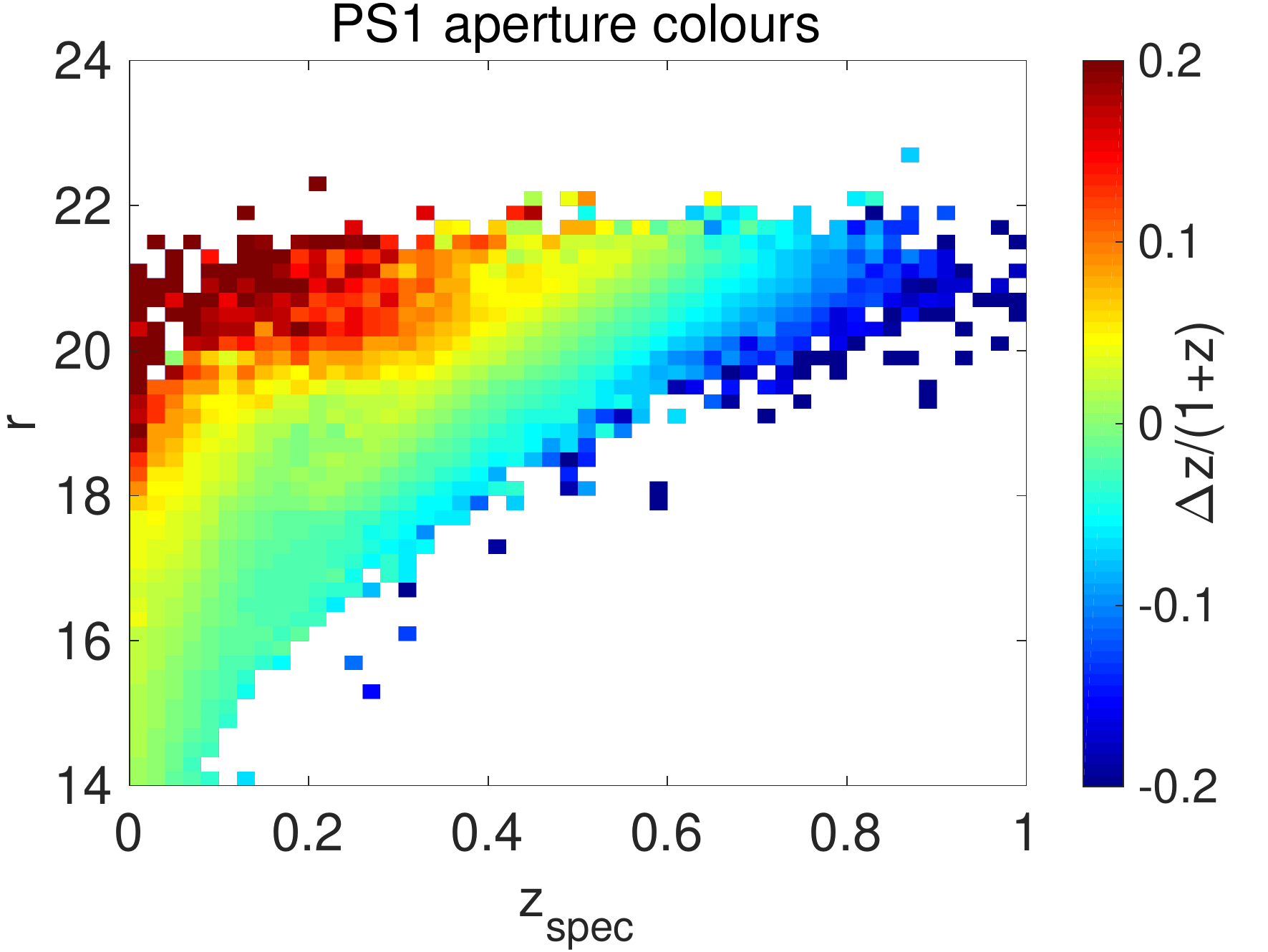}}
	\subfigure{\includegraphics[width=0.66\columnwidth]{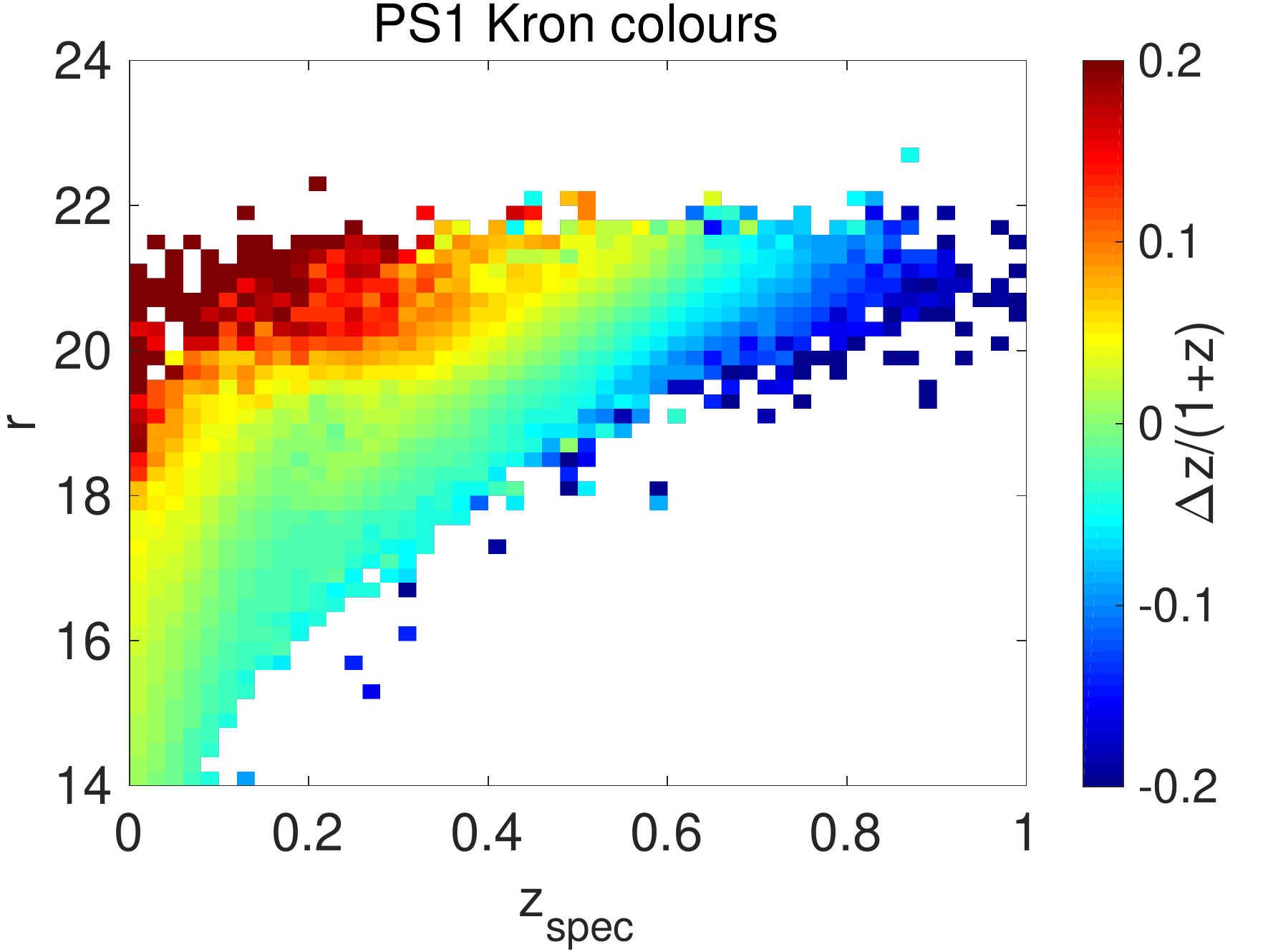}}
	\subfigure{\includegraphics[width=0.66\columnwidth]{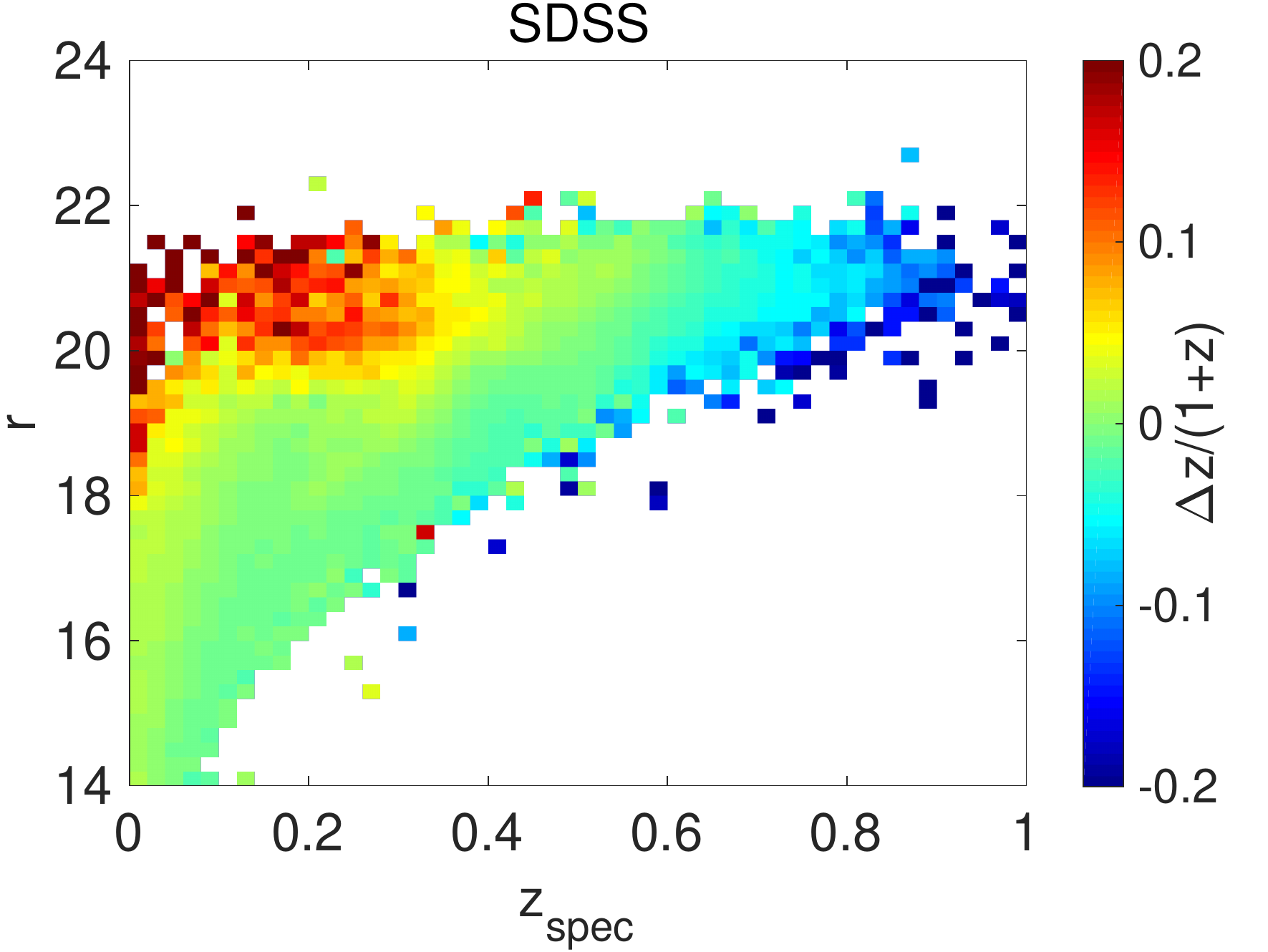}}
	\caption{Average normalized error $\Delta z_{\rm norm} = (z_{\rm phot}-z_{\rm spec})/(1+z_{\rm spec})$ in the 10 different $\mathcal{T}_{\rm test}$ sets as a function of the magnitude $r$ and the spectroscopic redshift $z_{\rm spec}$, for three different sets of features: left panel corresponds to PS1 features with aperture colours; middle panel corresponds to PS1 features with Kron colours; and right panel corresponds to SDSS features. In the three cases the training set is $\mathcal{T}_9^{\rm Beck}$.}
	\label{fig:SDSS_comparison}
\end{figure*}

\section{Comparison with other photometric features}\label{sec:feature_comparison}
The photometric redshift estimation method described in Sect. \ref{sec:method} is a general technique that can be applied to different sets of features. In the previous Section we presented the results obtained when using the five PS1 features described in Sect. \ref{ssec:feature_selection}, i.e., the PS1 $r$-band Kron magnitude and the $g-r$, $r-i$, $i-z$, and $z-y$ aperture colours. In this Section we analyse the effects of using different sets of features. In particular, we consider two different cases: 1) PS1 Kron colours, and 2) SDSS features, as in \cite{Beck2016}. 
In the first case, we will assess whether the Kron colors, which are not physically motivated (see Sect. \ref{ssec:feature_selection}) but  directly available for download for the complete PS1 survey, can be used if more convenient. Considering the SDSS features will allow us to compare the performance of the method using PS1 information with respect to the original method of Beck et al. (2016).

In order to do a fair comparison, we restricted our training set to the galaxies in $\mathcal{T}$ that have the $r$-band Kron magnitude, the four aperture colours, and the four Kron colours available in the PS1 dataset. Moreover, we also discard the galaxies that do not satisfy the colour cut and photometric error criteria defined in Eq. 7 of \cite{Beck2016}, based on SDSS information. The resulting training set, $\mathcal{T}_{9}^{\rm Beck}$, contains 1\,776\,508 galaxies. As in the previous experiments described in Sect. \ref{sec:results}, we divided 10 times $\mathcal{T}_{9}^{\rm Beck}$ into a training subset and a test subset. Then, we computed the photometric redshift of the galaxies in the 10 test sets using as features: 1) the  $r$-band Kron magnitude and the four aperture colours from PS1; 2) the $r$-band Kron magnitude and the four Kron colours from PS1; and 3) the five SDSS features defined in \cite{Beck2016}.

\begin{table}
	\caption{Average normalized redshift estimation bias $\overline{\Delta z_{\rm norm}}$, standard deviation $\sigma(\Delta z_{\rm norm})$ and outlier rate $P_o$ obtained when using different sets of features in the training set $\mathcal{T}_{9}^{\rm Beck}$. These quantities were calculated after iteratively removing the outliers, defined as $|\Delta z_{\rm norm}|>3\sigma(\Delta z_{\rm norm})$. }
	\label{table:errors_comparison}
	\centering 
	\begin{tabular}{c | c c c c}
		\hline
		\noalign{\smallskip}
		Features &  $\overline{\Delta z_{\rm norm}}$ &  $\sigma(\Delta z_{\rm norm})$ &    $P_o$  \\
		\noalign{\smallskip}
		\hline
		\noalign{\smallskip}
		PS1 aperture colours &     $4.70 \times 10^{-5}$ &  0.0280 &    3.09      \\
		PS1 Kron colours &          $1.88 \times 10^{-4}$&  0.0284 &    3.12     \\
		SDSS &        $-1.49 \times 10^{-4}$&  0.0198 &    3.83 \\ 
		\noalign{\smallskip}
		\hline
	\end{tabular}
\end{table}

Table \ref{table:errors_comparison} reports the average bias, standard deviation, and outlier rate obtained in the three cases. Figure \ref{fig:SDSS_comparison} shows the average normalized bias in the three cases as a function of the $r$-band Kron magnitude and the spectroscopic redshift $z_{\rm spec}$. The results obtained with PS1 aperture colours are not exactly the same as in the experiment presented in Sect. \ref{sec:results} (see Table \ref{table:errors} and Fig. \ref{fig:effect_of_r}) because the training set ($\mathcal{T}_{9}^{\rm Beck}$ instead of $\mathcal{T}_{5}$) contains now less galaxies. The galaxies that have been removed with respect to $\mathcal{T}_{5}$ are those for which a PS1 Kron magnitude is missing or that do not satisfy the SDSS criteria defined in \cite{Beck2016}. These are probably galaxies with poorer photometry, which explains the slight improvement in the performance (lower standard deviation and outlier rate) with respect to the results presented in Sect. \ref{sec:results}.

\subsection{Aperture colours vs Kron colours}
As shown in Table \ref{table:errors_comparison} and Fig. \ref{fig:SDSS_comparison}, the results obtained when using Kron colours are very similar to the ones obtained when using  aperture colours. We have also seen a nearly identical performance to the one shown in Figs. \ref{fig:zphot_zspec} and \ref{fig:histogram}: the photometric redshift calculated from Kron colours follows quite well the spectroscopic redshift, especially in the intermediate redshift range ($0.1<z_{\rm spec}<0.6$), and the method tends to underestimate the redshift for high redshift galaxies ($z_{\rm spec}>0.6$) and to overestimate the redshift for low redshift galaxies ($z_{\rm spec}<0.1$). 

Since the difference between using PS1 aperture colours  
or PS1 Kron colours is negligible, we have included the possibility of selecting which features to use in the code available at \url{www.testaddress.com}. The user may choose the one that is more convenient, without significant impact on the results.

\begin{figure*}
	\centering
	\subfigure{\includegraphics[width=0.99\columnwidth]{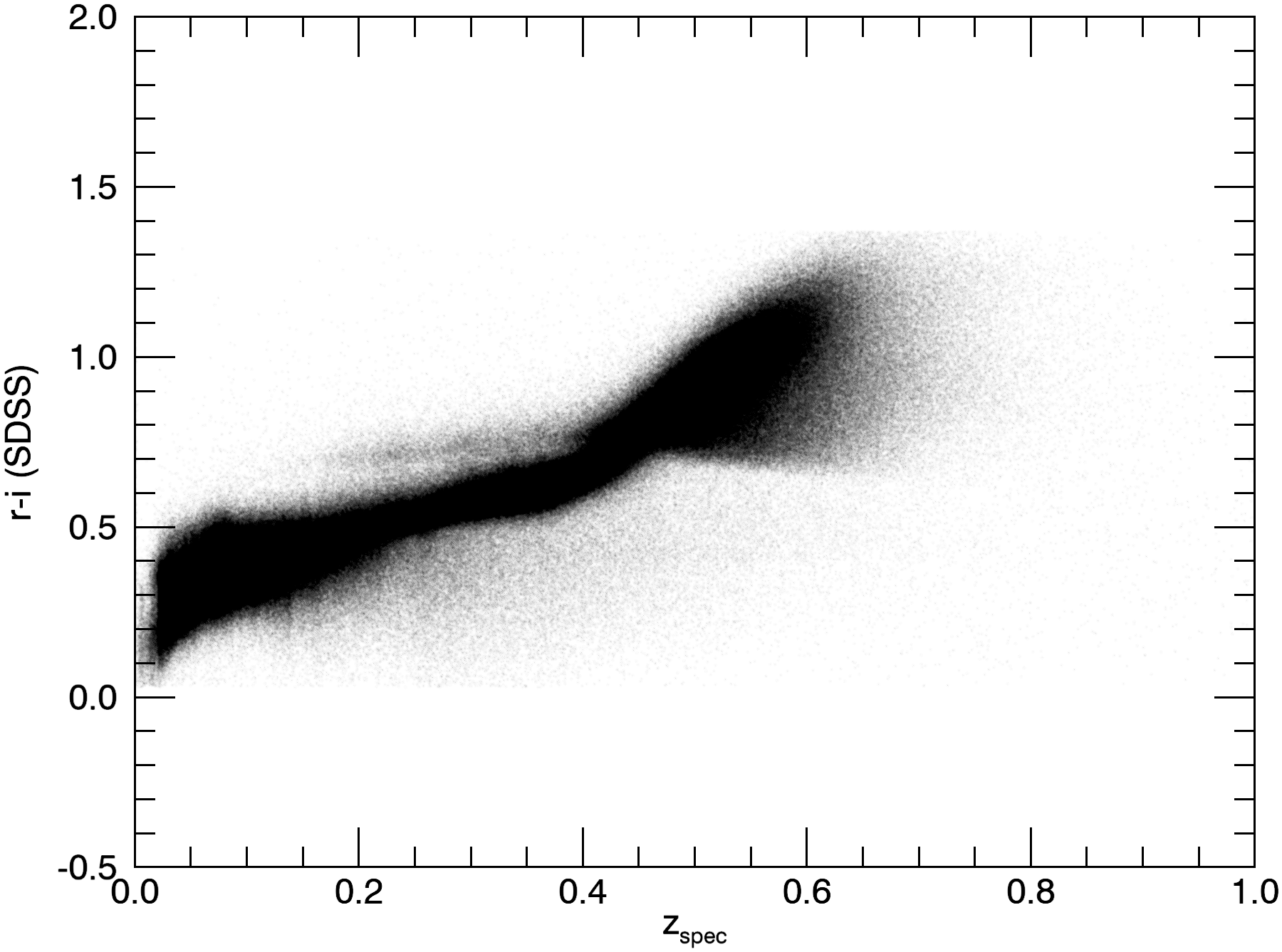}}
	\subfigure{\includegraphics[width=0.99\columnwidth]{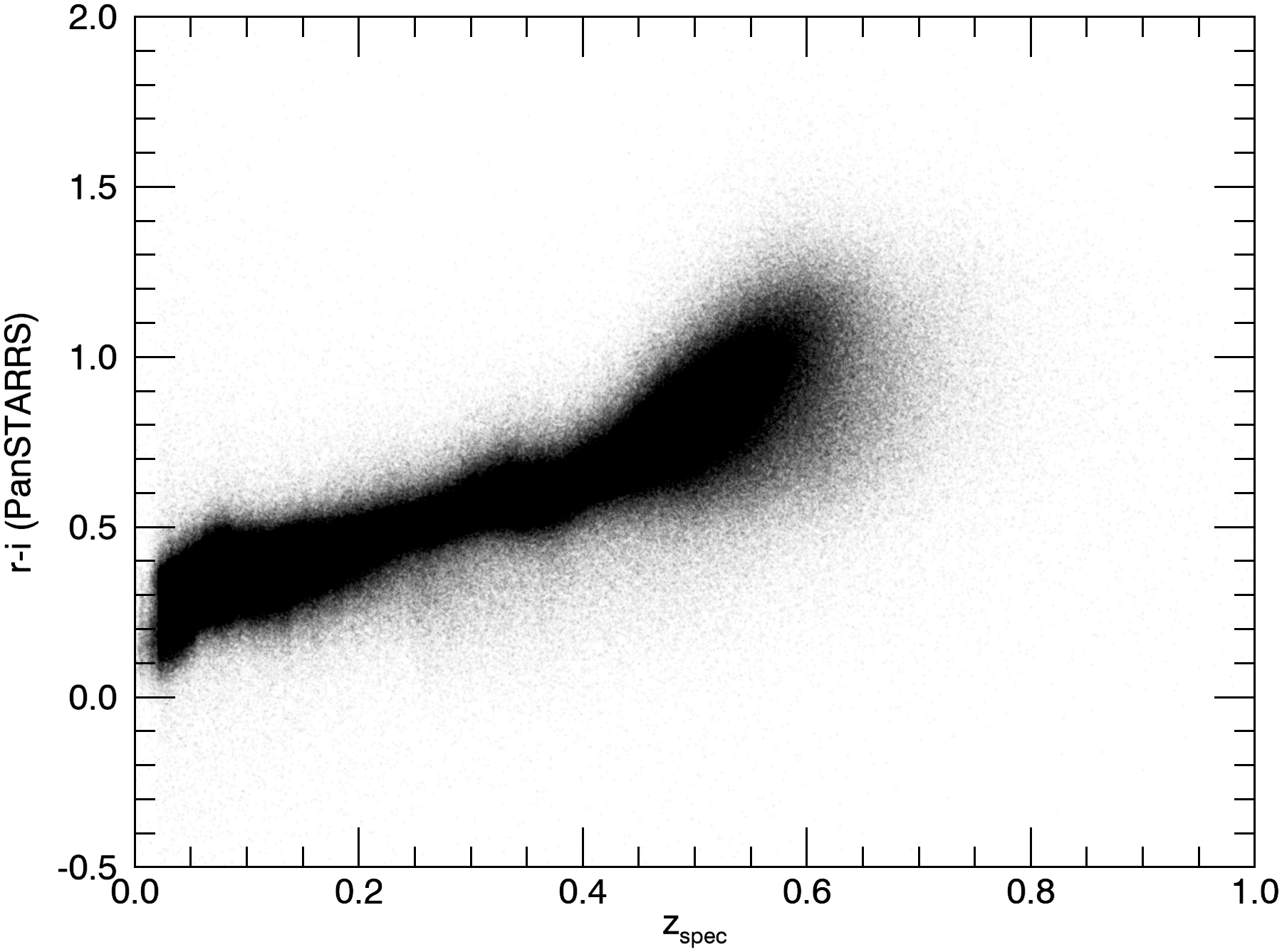}}
	\caption{Correlation between the $r-i$ colour and the spectroscopic redshift $z_{\rm spec}$ for SDSS (left panel) and PS1 (right panel) datasets. In the PS1 case, the aperture colour is represented.  Each point represents a galaxy of the training set $\mathcal{T}_9^{\rm Beck}$.}
	\label{fig:colour_vs_zspec}
\end{figure*}

\subsection{PS1 features vs SDSS features}\label{ssec:SDSS}
Figure \ref{fig:SDSS_comparison} and Table \ref{table:errors_comparison} show that using SDSS information instead of PS1 information results in a slightly better performance. The standard deviation of the normalized redshift error is lower with SDSS features. 
Moreover, although the global average bias is smaller for PS1 (with aperture colours), Fig. \ref{fig:SDSS_comparison} shows that SDSS features result in a lower bias both at high and low redshift. This effect is especially noticeable for fainter galaxies. In the following, we analyse the possible causes of this behaviour.

Firstly, PS1 and SDSS features are defined differently. One of the SDSS features is the $u-g$ colour, which is not available in Pan-STARSS; whereas the $z-y$ colour is available in PS1 but not in SDSS. Including a bluer information in SDSS allows a better estimation of the redshift of lower redshift galaxies. To check if this is enough to explain the observed behaviour, we repeated the calculation of the photometric redshift removing the $u-g$ colour from SDSS features and removing the $z-y$ colour from PS1 features. The results with SDSS features degrade, but still show a slightly better performance than with PS1 features, so this feature difference does not entirely explain the better behaviour of SDSS features.

Secondly, the magnitudes $g$, $r$, $i$ and $z$ have different values in SDSS and PS1. It turns out that SDSS magnitudes show a better correlation with the spectroscopic redshift, which explains their power to better estimate the photometric redshift. Figure \ref{fig:colour_vs_zspec} shows a comparison of the correlation between the $r-i$ colour and the spectroscopic redshift for SDSS and PS1 (considering the $r-i$ aperture colour). For a given redshift, PS1 values show a larger dispersion than SDSS. The same occurs for the $g-r$ and $i-z$ aperture colours, for the Kron colours, and for the $r$ Kron magnitude. The reason for this larger scatter could be that PS1 aperture colours are not measured exactly at the Kron radius, but at the closest one from the five available apertures, resulting in colours that do not correspond to the same percentage of flux for all the galaxies. Conversely, SDSS colours are computed from {\tt ModelMag} magnitudes, so they correspond to the total flux of the galaxy. On the other hand, PS1 Kron colours are not physically motivated, since they are calculated as the difference between two magnitudes that may be measured in different radii.

\begin{figure*}
	\centering
	\subfigure{\includegraphics[width=0.66\columnwidth]{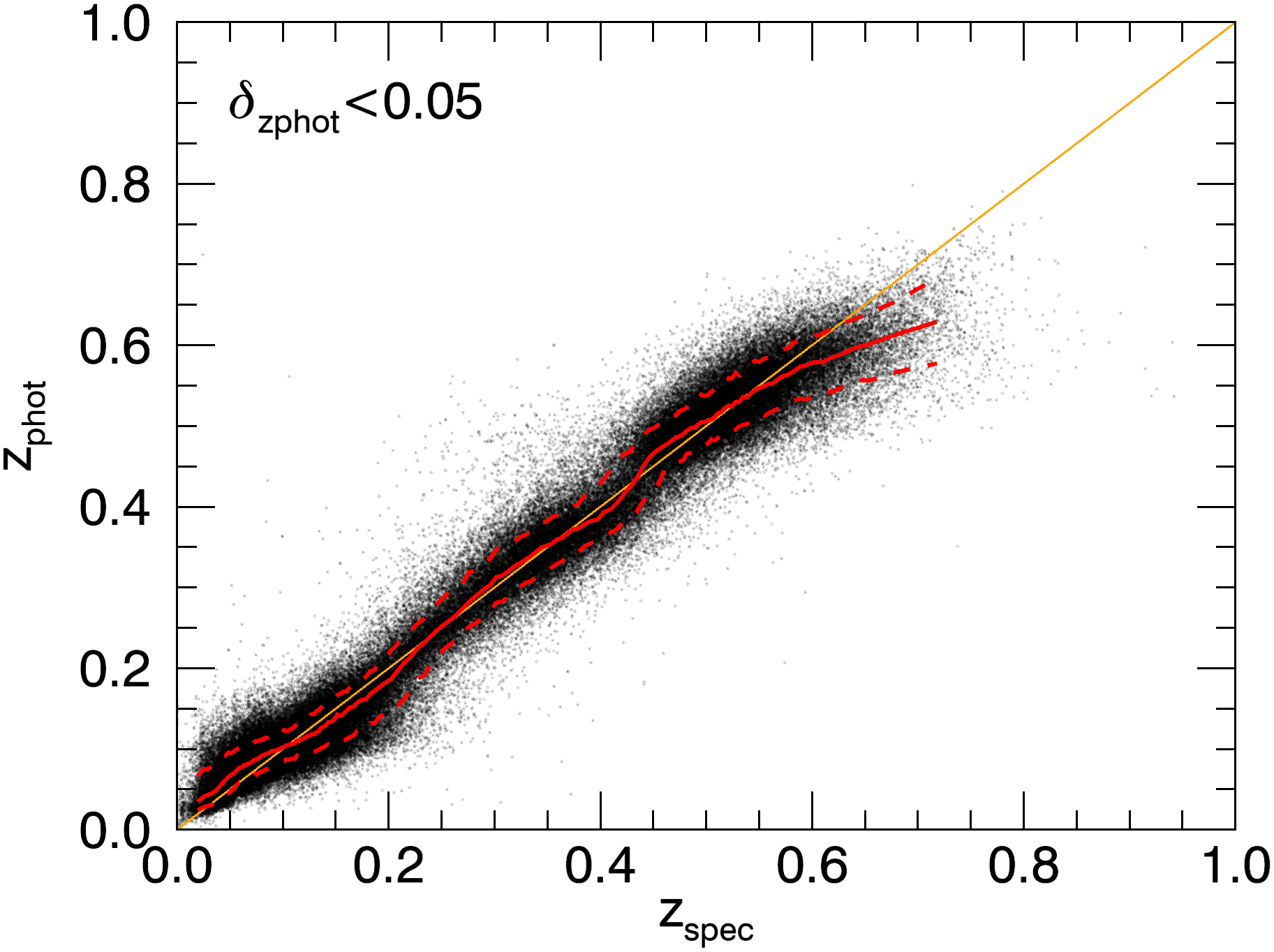}}
	\subfigure{\includegraphics[width=0.66\columnwidth]{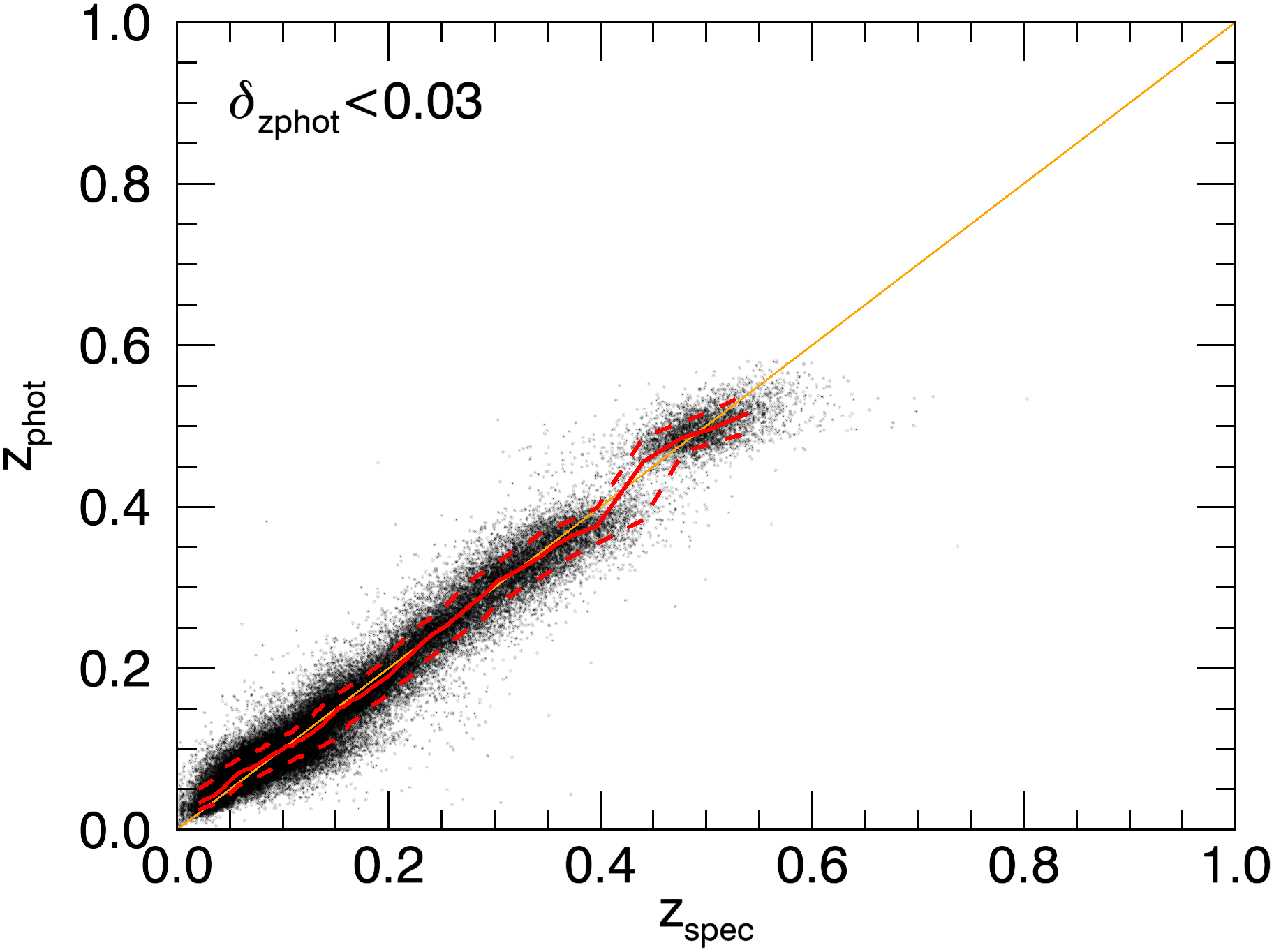}}
	\subfigure{\includegraphics[width=0.66\columnwidth]{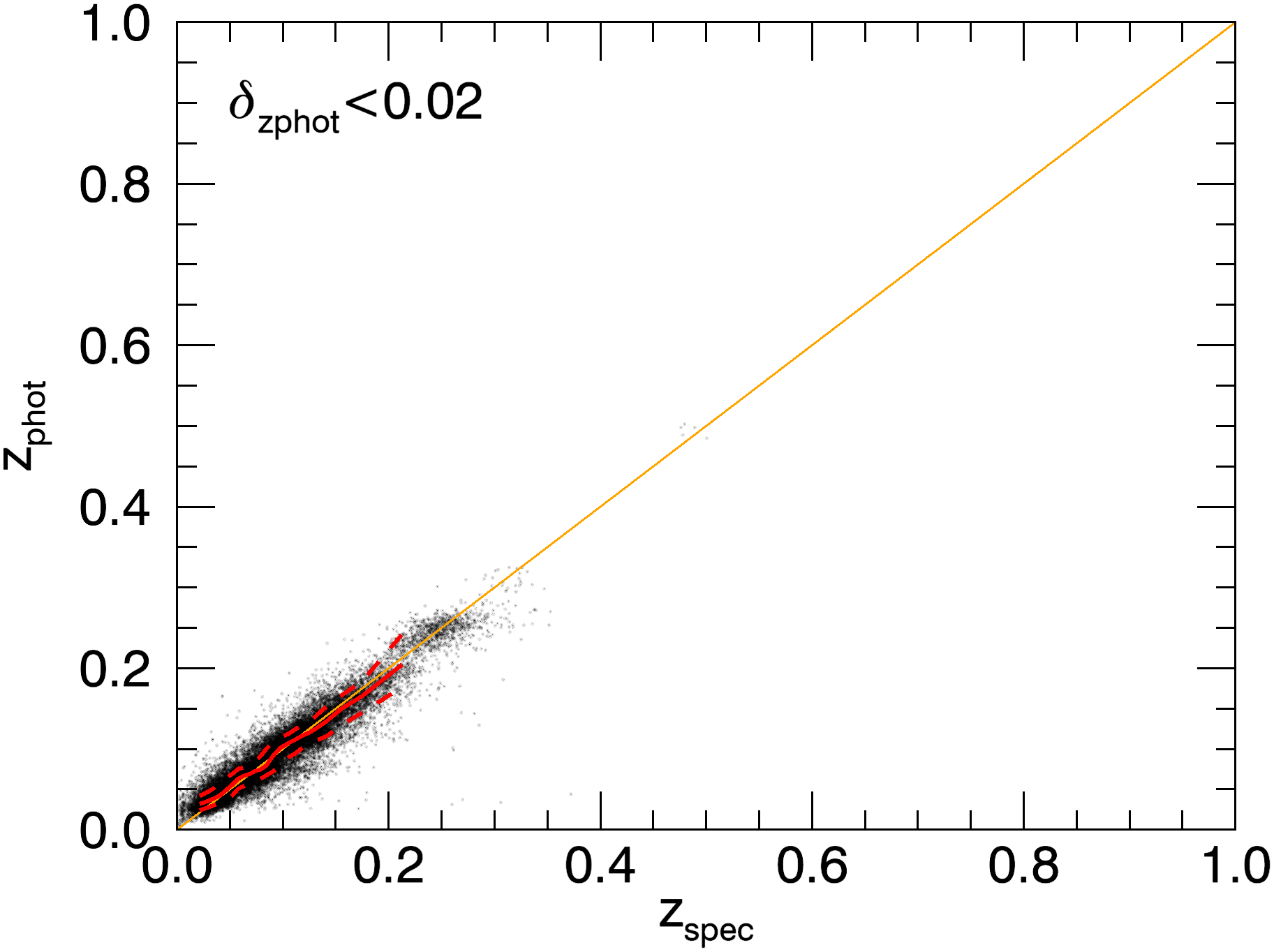}}
	\caption{Photometric redshift as a function of spectroscopic redshift, for three different sets. The red solid and dashed lines represent the median and the 68\% confidence regions, respectively, computed in small $z_{\rm spec}$ intervals. The orange line shows $z_{\rm phot}=z_{\rm spec}$. On panel (a), we included the galaxies  with a reported redshift error of $\delta_{z_{\rm phot}}<0.05$. On panel (b), we included the galaxies with a reported redshift error of $\delta_{z_{\rm phot}}<0.03$. On panel (c), we included the galaxies with a reported redshift error of $\delta_{z_{\rm phot}}<0.02$. }
	\label{fig:different_cuts}
\end{figure*}

\begin{figure*}
	\centering
	\subfigure{\includegraphics[width=0.66\columnwidth]{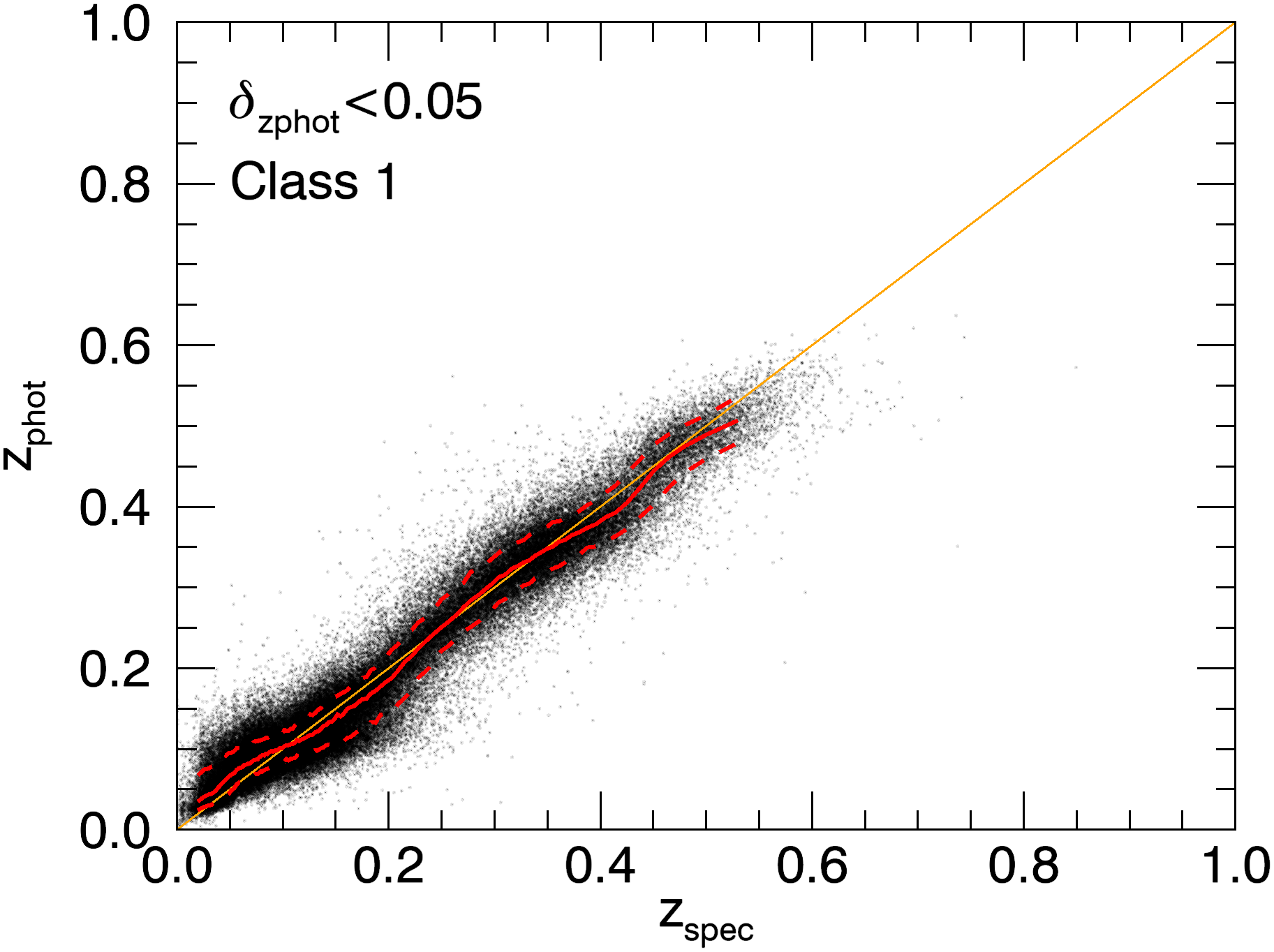}}
	\subfigure{\includegraphics[width=0.66\columnwidth]{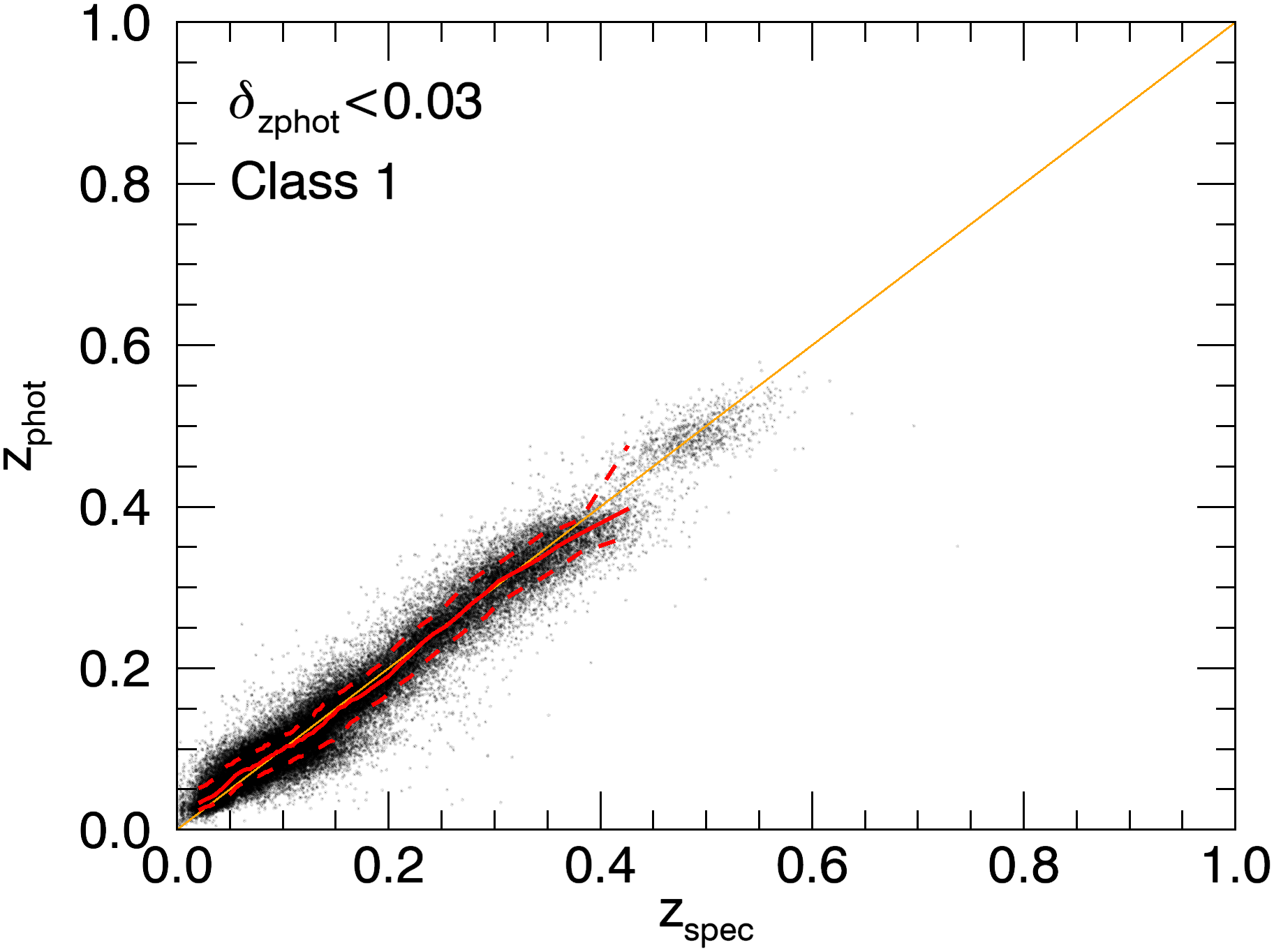}}
	\subfigure{\includegraphics[width=0.66\columnwidth]{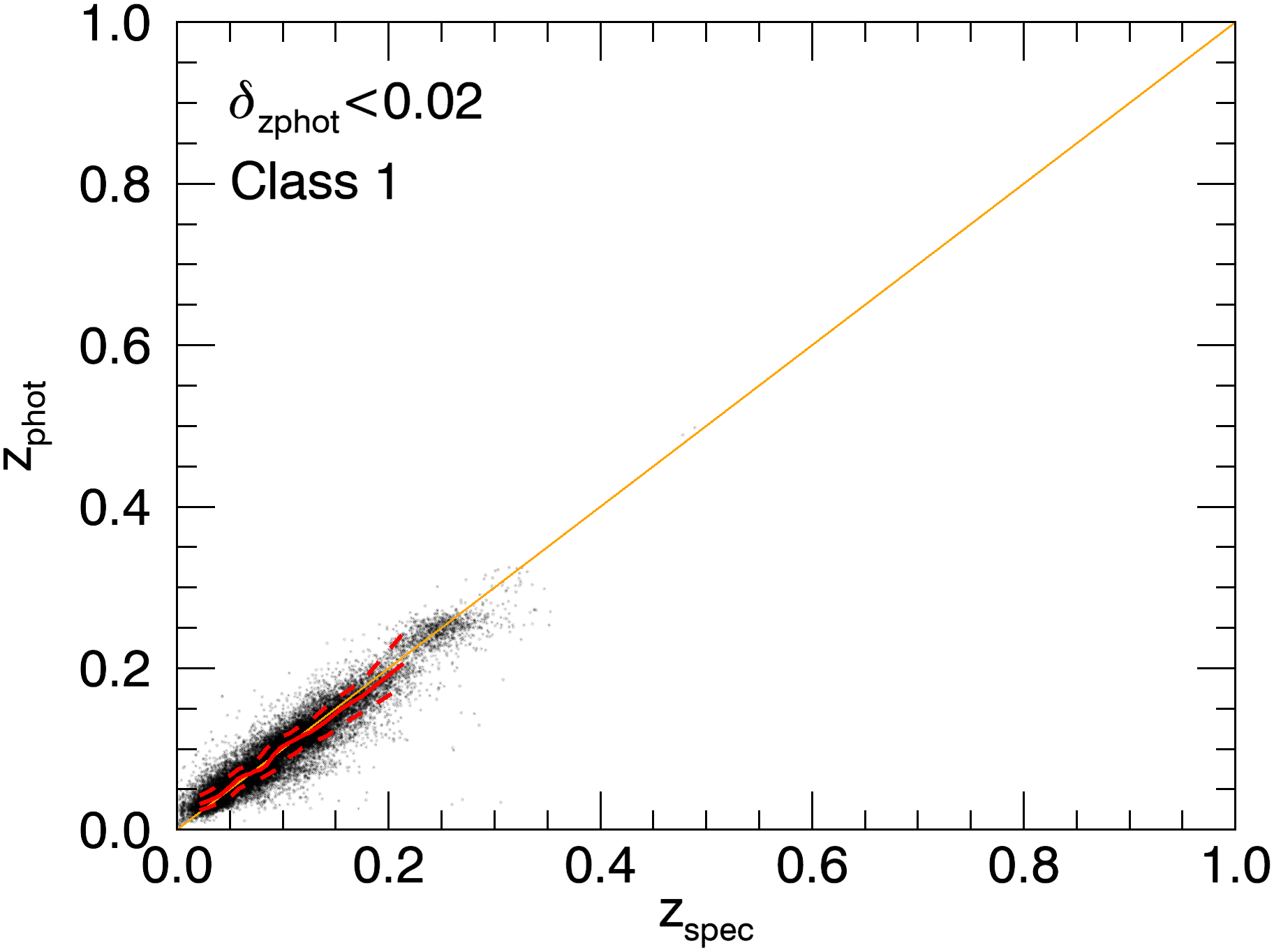}}
	\caption{Photometric redshift as a function of spectroscopic redshift, for three different sets. The red solid and dashed lines represent the median and the 68\% confidence regions, respectively, computed in small $z_{\rm spec}$ intervals. The orange line shows $z_{\rm phot}=z_{\rm spec}$. On panel (a), we included the galaxies in photometric error class 1 and with a reported redshift error of $\delta_{z_{\rm phot}}<0.05$. On panel (b), we included the galaxies in photometric error class 1 and with a reported redshift error of $\delta_{z_{\rm phot}}<0.03$. On panel (c), we included the galaxies in photometric error class 1 and with a reported redshift error of $\delta_{z_{\rm phot}}<0.02$. }
	\label{fig:different_cuts2}
\end{figure*}

\section{Practical guidelines for using the method}\label{sec:best_practices}

The training set $\mathcal{T}$ and the code implementing our method are available for download at the following webpage: \url{www.testaddress.com}. This allows the estimation of the photometric redshift of any galaxy in the PS1 survey. The code includes several configuration options that are described in detail in the webpage. The two main options are the choice between aperture or Kron magnitudes, and the choice of the subset of $\mathcal{T}$ to be used for training. 

Depending on the required accuracy and on the specific use of the photometric redshifts provided by our method, one may want to use all the possible photometric redshifts regardless of their error, or prefer to use a lower amount of more accurate photometric redshifts. In this Section, we summarize the different ways that we provide of selecting the best redshifts.

There are three main parameters that can be used for this selection: the estimated error $\delta_{z_{\rm phot}}$, the photometric error class, and the extrapolation flag. Figure \ref{fig:different_cuts} shows the effect of using different cuts in the photometric redshift errors. As expected, introducing a cut in $\delta_{z_{\rm phot}}$ reduces the errors (see for comparison Fig. \ref{fig:zphot_zspec}, where no cuts were used). However, if the cut is too severe, the resulting sample may be limited in terms of redshift and colour space coverage. Therefore, we suggest to test different values to find the most appropriate for a particular goal. Figure \ref{fig:different_cuts2} shows the effect of adding a cut using the photometric error class. By comparing Fig. \ref{fig:different_cuts} and Fig. \ref{fig:different_cuts2} we can see that the photometric error class selection mainly reduces the scatter at high redshifts, where the photometric errors are larger. 

Given the small number of galaxies in our dataset for which an extrapolation was performed, filtering them out does not bring a noticeable effect on the ensemble. However, this parameter is a good indicator of the accuracy of the results, as shown in Table \ref{table:errors}, so it can be used to filter out some of the calculated redshifts when there is a need for high accuracy.

Finally, the error maps presented in Sect. \ref{ssec:error_maps} can be used to filter out some of the galaxies located in the regions of the magnitude-colour space that are more prone to errors.

\section{Summary}\label{sec:summary}
We present a data-driven method to compute photometric redshifts for galaxies using the PS1 survey. In this work we use data from the PS1 DR2, but we tested the results also for the DR1, finding no significant difference. 
Our method is an adaptation of the one proposed by \citet{Beck2016} for the SDSS DR12, based on a local linear regression in a 5-dimensional magnitude and colour space. To adapt \cite{Beck2016} algorithm to PS1 we select appropriate magnitudes and colours ($r$, $g-r$, $r-i$, $i-z$, and $z-y$) for defining the 5-dimensional space, and we construct a proper and clean training set composed of 2\,313\,724 galaxies, whose spectroscopic redshift is available from SDSS and whose magnitudes and colours are obtained from the PS1 DR2 survey.

We assess the performance of this method by means of a cross-validation on the training set, i.e., we use part of the galaxies of our training set as test galaxies to estimate their photometric redshifts and we then compare them to their true (spectroscopic) redshifts. We estimate that the average bias of our method is $\overline{\Delta z_{\rm norm}}=-2.01 \times 10^{-4}$, its standard deviation, $\sigma(\Delta z_{\rm norm})=0.0298$, and the outlier rate $P_o=4.32\%$. 

We also evaluate the impact of the photometric uncertainties on our redshift determination. This was done by dividing the entire sample in 5 photometric classes of growing photometric errors. As expected, the uncertainties on the photometric redshifts are smaller where the photometric errors are smaller. 
There is also a fraction of galaxies for which the method extrapolates the photometric redshift, since their features lie outside the bounding box of their nearest neighbours. In these cases, the errors on photometric redshifts are larger and these galaxies are flagged appropriately.

Moreover, we analyse the impact of the galaxy density (in the feature space) on the redshift determination. In fact, there are regions in the 5-dimensional space that are more populated than others. As expected, we find that galaxies located in crowded regions have a better redshift estimation than galaxies found in sparse regions. 

Since galaxies in PS1 may have incomplete photometry, our method is prepared to deal with the case of missing features. We evaluate the effect that a missing feature may produce on the results using an ablation test (artficially removing existing features). As expected, the scatter increases in these cases. This effect is especially important when the $r$ Kron magnitude is missing, whereas a missing colour has a smaller impact.

Furthermore, we test the use of PS1 Kron colours instead of aperture colours, finding no significant difference in the results. Although Kron colours have no physical meaning, they are easier to obtain and it is worth stressing that they can be safely used for computing photometric redshifts with our method.

Finally, we compare our results with those presented in \citet{Beck2016} for the SDSS DR12. We find that SDSS data perform slightly better than PS1 features, especially for faint galaxies ($r > 20$). We suggest that these differences could be caused by two main factors: different available filters, with SDSS offering a bluer band, and a stronger correlation between the SDSS magnitudes and the spectroscopic redshift. 
However, it is worth noticing that the overall performance of our method is fully in agreement, within the uncertainties, with the SDSS results. 

A version of the code and training set is available for download at the following webpage: \url{www.testaddress.com}.

\begin{acknowledgements}
	The research leading to these results has received funding from the European Research Council under the European Union's Seventh Framework Programme (FP7/2007-2013) / ERC grant agreement n$^{\circ}$ 340519.
	
	The authors would like to thank Monique Arnaud for helpful discussions and suggestions.
	
	This research has made use of the Pan-STARRS1 Survey. The Pan-STARRS1 Surveys (PS1) and the PS1 public science archive have been made possible through contributions by the Institute for Astronomy, the University of Hawaii, the Pan-STARRS Project Office, the Max-Planck Society and its participating institutes, the Max Planck Institute for Astronomy, Heidelberg and the Max Planck Institute for Extraterrestrial Physics, Garching, The Johns Hopkins University, Durham University, the University of Edinburgh, the Queen's University Belfast, the Harvard-Smithsonian Center for Astrophysics, the Las Cumbres Observatory Global Telescope Network Incorporated, the National Central University of Taiwan, the Space Telescope Science Institute, the National Aeronautics and Space Administration under Grant No. NNX08AR22G issued through the Planetary Science Division of the NASA Science Mission Directorate, the National Science Foundation Grant No. AST-1238877, the University of Maryland, Eotvos Lorand University (ELTE), the Los Alamos National Laboratory, and the Gordon and Betty Moore Foundation.
	
	This research has also made use of the SDSS-III Survey (DR12). Funding for SDSS-III has been provided by the Alfred P. Sloan Foundation, the Participating Institutions, the National Science Foundation, and the U.S. Department of Energy Office of Science. The SDSS-III web site is http://www.sdss3.org/.  
	
	Some of the data presented in this paper were obtained from the Mikulski Archive for Space Telescopes (MAST). STScI is operated by the Association of Universities for Research in Astronomy, Inc., under NASA contract NAS5-26555. Support for MAST for non-HST data is provided by the NASA Office of Space Science via grant NNX13AC07G and by other grants and contracts. 
	
	The authors acknowledge the use of the Catalog Archive Server Jobs System (CasJobs) service at http://casjobs.sdss.org/CasJobs, developed by the JHU/SDSS team.
 
\end{acknowledgements}

%
%
\bibliographystyle{aa} 
\bibliography{bibliografia} 

\end{document}